\documentclass[12pt]{article}
\usepackage{amsmath,amssymb,amsthm}
\usepackage{graphicx}
\usepackage{natbib}
\usepackage{url} 
\usepackage{tikz}
\usepackage{color}
\usetikzlibrary{calc,patterns,quotes}
\usetikzlibrary{shapes.misc}

\tikzset{cross/.style={cross out, draw=black, minimum size=2*(#1-\pgflinewidth), inner sep=0pt, outer sep=0pt},
cross/.default={1pt}}

\DeclareMathOperator{\cv}{CV}
\DeclareMathOperator{\df}{df}
\DeclareMathOperator{\tr}{tr}
\DeclareMathOperator{\wCV}{wCV}

\newtheorem{lemma}{Lemma}

\newtheorem{definition}{Definition}

\theoremstyle{remark}

\newcommand{\vc}[1]{\mathbf{#1}}
\newcommand{\mat}[1]{\mathbf{#1}}

\begin{document}

 \title{\bf Influence of single observations on the choice of the penalty parameter in ridge regression}
  \author{Kristoffer H. Hellton\thanks{hellton@nr.no} \\
    Norwegian Computing Center, Norway\\\\
    Camilla Lingj\ae rde \\
Norwegian Computing Center, Norway \\
MRC Biostatistics Unit, Cambridge University, UK 
\\
\\
Riccardo De Bin
\\
Department of Mathematics, University of Oslo, Norway
}
\maketitle

\begin{abstract}
Penalized regression methods such as ridge regression heavily rely on the choice of a tuning or penalty parameter, which is often computed via cross-validation. Discrepancies in the value of the penalty parameter may lead to substantial differences in regression coefficient estimates and predictions. In this paper, we investigate the effect of single observations on the optimal choice of the tuning parameter, showing how the presence of influential points can change it dramatically. We distinguish between points as ``expanders'' and ``shrinkers'', based on their effect on the model complexity. Our approach supplies a visual exploratory tool to identify influential points, naturally implementable for high-dimensional data where traditional approaches usually fail. Applications to simulated and real data examples, both low- and high-dimensional, are presented. The visual tool is implemented in the R package \emph{influridge}. 
\end{abstract}

\noindent%
{\it Keywords:} Genomic data; High dimensional data; Influential points; Leverage points; Outliers; Penalized regression.

\section{Introduction}
Model instability is a well-known problem in statistics. It refers to the phenomenon for which small changes in the data cause large differences in the final statistical model \citep{Breiman1996, HeinzeAl2018}. For example, if one repeatedly applies a variable selection procedure, such as backward elimination, on slightly perturbed sets of data, many different models can be obtained \citep{DebinAl2015}. Even when the same model is selected, or no selection is implemented, noticeable differences in terms of coefficient estimates may occur: adding or removing observations in the dataset normally modify the final estimates, especially in the case of low sample sizes. When the difference due to the effect of a single observation is substantial, that observation is defined as an \emph{influential point} \citep{cook1979influential}. In the low dimensional setting ($p < n$, where $p$ is the number of variables, $n$ the sample size), several methods to evaluate the effect of single observations on the coefficient estimates \citep[see, e.g.,][]{cook1979influential, BelsleyAl1980, Penha2005} or on the selection of the variables \citep[e.g.,][]{AtkinsonRiani2002, DeBin17Influential} have been developed, eventually leading to approaches to identify influential points. Over the last few years, some work has been done in the high-dimensional setting ($p > n$) as well, including \cite{ZhaoAl2013, ZhaoAl2019}, \cite{WangLi2017}, \cite{WangAl2018} and \cite{RajaratnamAl2019}. These methods basically adapt traditional low-dimensional tools to work in the high-dimensional setting: for example, \cite{ZhaoAl2013} and \cite{WangLi2017} extended the \cite{cook1979influential}'s distance, shifting the attention from the observation's influence on the least squares regression estimates to the influence on the marginal correlations and on the distance correlations, respectively. \cite{RajaratnamAl2019} started from the idea behind the DFBETA measure \citep{BelsleyAl1980} to develop four influence measures applied to the lasso, focusing on the difference in the number of selected variables between using all data and deleting an observation (termed \emph{df-model}). Another measure they introduced, \emph{df-lambda}, quantified the change in the optimal value of the lasso tuning parameter when removing one observation.

Focusing on ridge regression, the same strategy has been pursued by \cite{WalkerBirch1988}: in this case, they adapted the DFFITS measure \citep{BelsleyAl1980}. Based on this work, \cite{ShiWang1999} investigated the influence of single observations on the choice of the penalty parameter in ridge regression. While clearly connected to ours, their work considers a specific, and not widely used, procedure to find the tuning parameter, the minimization of \cite{Myers1986}'s $C_\lambda$. With this noticeable exception, not much attention has been devoted to investigating the influence of a single observation on the choice of the penalty parameter in a regularized regression setting. It is known that in this framework different values of this parameter lead to very different models \citep[see, e.g.,][]{MeinshausenBuehlmann2010}, hence it is highly important to understand the role of each observation on that. In particular, we show how specific observations push towards more complex models by requiring a smaller value of the tuning parameter (in the following, \emph{expanders}), while others have the opposite effect (\emph{shrinkers}). Points with extreme effects can be recognized as influential points.

In contrast to the aforementioned work of \cite{ShiWang1999}, we consider a procedure based on cross-validation. In particular, we exploit the leave-one-out cross-validation, that allows us to easily investigate the particular influence of a single observation. Our procedure, nevertheless, can be used regardless of the method used to select the tuning parameter itself. In this paper we focus on the tuning parameter of ridge regression, but in principle our approach works for any regularized regression technique (see Section \ref{sect:discussion} for more details). Following the tradition of differentiation methods \citep{BelsleyAl1980}, we investigate how the choice of the tuning parameter evolves when the weight assigned to a specific observation is perturbed. This allows us to have a better grasp of the overall influence of this observation, not limited to the dichotomy presence/absence typical of deletion approaches \citep[see][for a discussion on differentiation and deletion methods]{BelsleyAl1980, RajaratnamAl2019}. We propose to visualize the observation's influence through a curve and study its behavior to characterize its effect on the choice of the tuning parameter. Using graphical tools to investigate the influence of observations was, for instance, utilized by \citet{genton2010visualizing} for dependent data from time series analysis and spatial statistics. 

Other recent work studying the influence of observations in a high-dimensional setting, include \citet{yatracos2022residual}, who introduce the Residual's influence index (RINFIN) depending on partial derivatives of the regression coefficients' influence functions, and \citet{barry2022asymmetric}, using the concept of expectiles to develop an influence measure based on the asymmetric marginal correlation.

The paper is organized as follows: in Section \ref{sect:ridge} we briefly review ridge regression and the cross-validation approach for finding its tuning parameter. The influence of a single observation in this procedure is studied analytically in Section \ref{sec:InfluentialMethodology} and via simulation in Section \ref{sect:simulations}. Illustrative examples using real data, both in a low- and a high-dimensional setting, are shown in Section \ref{sect:realData}. A final discussion in Section \ref{sect:discussion} completes the paper.

\section{Ridge regression}\label{sect:ridge}
Ridge regression was originally introduced by \citet{hoerl1970ridge} to handle rank deficient data matrices and multicollinearity. The ridge (or $L_{2}$) penalty avoids these problems by ensuring the invertibility of the sample covariance matrix, shrinking the regression coefficients towards zero. The introduced penalty is controlled by a tuning parameter, $\lambda$, which requires data-dependent tuning. In the case of ridge regression, there exists a myriad of tuning procedures, but $K$-fold cross-validation has emerged as the standard within statistics, typically with  $K=10$ or $K=5$ \citep[after][]{hastie2001elements}. Following this approach, the data are randomly divided into $K$ parts (or folds), where each fold is held out and predicted by fitting the model on the remaining folds. The prediction error is averaged over all folds and computed for a range of tuning parameter values. The value leading to the lowest error is then selected. A special case, important for ridge regression, is the leave-one-out cross-validation, or $n$-fold cross-validation, in which each observation constitutes a separate fold, requiring no random division. 

Leave-one-out cross-validation is particularly relevant for ridge regression because an explicit expression for the error can be derived \citep[see for instance][]{golub1979generalized}. Consider a linear regression model where we have observed $n$ univariate continuous outcomes, $y_{i}\in\mathbb{R}$, and $p$-dimensional covariate vectors, $\vc{x}_{i} \in \mathbb{R}^p$, 
\begin{equation*}
y_{i}= \vc{x}_{i}^T\beta+ \varepsilon_{i}, \quad i=1,\dots,n.
\end{equation*}
Here $\beta\in \mathbb{R}^p$ is a $p$-dimensional vector of regression coefficients and $\varepsilon_i \in \mathbb{R}$ are identically and independently distributed noise terms with zero mean. For an $n \times p$ data matrix $\mat{X}$ and a vector of outcomes $\vc{y} = [y_1,\dots,y_n]^T$, the predictions of ridge regression for a fixed $\lambda$ are given by
\begin{equation*}
\hat{\vc{y}}(\lambda)  =  \mat{X} \hat{\vc{\beta}}(\lambda) =  \mat{X} ( \mat{X}^T \mat{X}+ \lambda I_p)^{-1}  \mat{X}^T  \vc{y} =  \mat{H}(\lambda)  \vc{y},
\end{equation*}
where $\mat{H}(\lambda)$ is referred to as the hat matrix of ridge regression and $I_p$ is the identity matrix of dimension $p$. When we consider leave-one-out cross-validation, the regression coefficients for each fold and the prediction error for the removed observation are given by
\begin{equation*}
\hat{\beta}_{[i]}(\lambda) = (\mat{X}_{[i]}^T \mat{X}_{[i]}+\lambda I_{p})^{-1}\mat{X}_{[i]}^T \vc{y}_{[i]}, \quad \quad \text{ and } \quad\quad e_{[i]}(\lambda)= y_{i} - x_{i}^T\hat{{\beta}}_{[i]}(\lambda),
\end{equation*}
respectively. Here $\mat{X}_{[i]}$ and $\vc{y}_{[i]}$ are the data matrix and outcome vector excluding the $i$th row. Note that we denote the prediction error for the $i$th observation as $e_{[i]}$ to distinguish it from the residual $e_i(\lambda) = y_{i} - x_{i}^T\hat{{\beta}}(\lambda)$, in which $\beta$ is estimated using all observations. \citet{golub1979generalized} stated (see Section S.1 in the Supplementary material for a detailed derivation) that the leave-one-out cross-validation error of ridge regression can be explicitly expressed as a function of the residuals $e_i$ by using the Sherman-Morrison-Woodbury formula for matrix inverses,
\begin{equation}\label{cv_res}
\cv(\lambda) = \frac{1}{n} \sum_{i=1}^n e_{[i]}(\lambda)^2 = \frac{1}{n} \sum_{i=1}^n \left(\frac{e_i}{1-H_{ii}(\lambda)}\right)^2,
\end{equation}
where $H_{ii}(\lambda)=\vc{x}_{i}^T(\mat{X}^T \mat{X}+\lambda I_{p})^{-1}\vc{x}_{i}$ is the $i$th element of the diagonal of $H(\lambda)$. The optimal cross-validation tuning parameter, $\lambda \geq 0$, is then the minimizer of the cross-validation error  
\begin{equation}\label{cv}
 \hat\lambda_{\cv} = \arg\min_{\lambda \geq 0 } \left\{\sum_{i=1}^n \left( \frac{y_{i} - \hat y_{i}(\lambda)}{1-H_{ii}(\lambda)} \right)^2\right\},
\end{equation}
where $\hat{y}_i(\lambda)$ is the $i$th element of $\hat{\vc{y}}(\lambda)$. Equation \eqref{cv} shows that the leave-one-out cross-validation tuning parameter $\hat\lambda_{\cv}$ minimizes a weighted version of the residuals. The weights are related to the \emph{leverage} of the data points through $H_{ii}(\lambda)$. The leverage measures how far away the covariates of an observation are from those of the other observations. Observations with high leverage have typically extreme or outlying covariate values and a lack of neighboring observations causes the fitted regression model to pass close to that particular observation. In contrast to the ordinary least square (hereafter, OLS) version, the hat matrix $H_{ii}(\lambda)$ also account for the effect of the penalization and therefore depends on the tuning parameter. 

\section{Influence of the single observations}\label{sec:InfluentialMethodology}

\subsection{Weighted cross-validation}\label{subsect:weightedCV}
To study how the observations influence the choice of the tuning parameter, we utilize a continuously weighted version of the leave-one-out cross-validation criterion, 
\begin{equation*}
\wCV(\lambda) =  \sum_{i=1}^n w_ie_{[i]}(\lambda)^2,
\end{equation*}
with normalized weights $w_1, \dots, w_n$ fulfilling $\sum_{i=1}^n w_i = 1$. Here uniform weights, $w_i = \frac{1}{n}$ for $i=1, \dots, n$, correspond to the standard leave-one-out cross-validation criterion. As $e_{[i]}(\lambda)$ gives the \emph{out-of-sample error} of the $i$th observation, changing the hypothetical weight of the observation in the dataset would not change the error $e_{[i]}(\lambda)$ itself (or the model used to estimate the error). A weight of zero, $w_{i}= 0$, corresponds to the exclusion of the $i$th residual, while weights larger than $1/n$ correspond to a ``continuous'' number of copies of the observation. The criterion further approximates a continuous version of the bootstrap approach of \citet{DeBin17Influential}, except for the fact that additional bootstrap copies of the observation would change the model and hence the prediction error $e_{[i]}(\lambda)$. 

We quantify the effect of a single observation on the optimal tuning parameter by varying the $i$th weight only. Under the normalization, all other weights are $w_j = \frac{1-w_{i}}{n-1}$ for $j=1,\dots,i-1,i+1,\dots,n$. 
\begin{definition}[Single normalized weight cross-validation] \label{eq:weightedCV}
The single normalized weight cross-validation criterion is defined as
\begin{equation}
\wCV(\lambda,w_i) = w_{i} e_{[i]}^2(\lambda) + \sum_{j \neq i } \frac{1-w_{i}}{n-1} e_{[j]}(\lambda)^2, \label{weightedCrit}
\end{equation} 
where $w_i\in[0,1]$ is the weight related to the $i$th observation. The word ``normalized'' here implies that $w_{i} + \sum_{j \neq i }^n \frac{1-w_{i}}{n-1} = 1$. For ridge regression, the criterion is then given by
\begin{equation*}
\wCV(\lambda,w_i) = w_{i} \left( \frac{y_{i} - \hat y_{i}(\lambda)}{1-H_{ii}(\lambda)} \right)^2+ \sum_{j \neq i } \frac{1-w_{i}}{n-1} \left( \frac{y_{j} - \hat y_{j}(\lambda)}{1-H_{jj}(\lambda)} \right)^2. 
\end{equation*}
\end{definition}
With the $i$th normalized weight $w_i$ ranging from 0 to 1, again $w_i=0$ corresponds to removing the impact of the residual of the $i$th observation with all other observations equally up-weighted, while the weight $w_i=1$ corresponds to removing all other residuals.

\subsection{Shrinkers and expanders}\label{subsect:shrinkersExpanders}
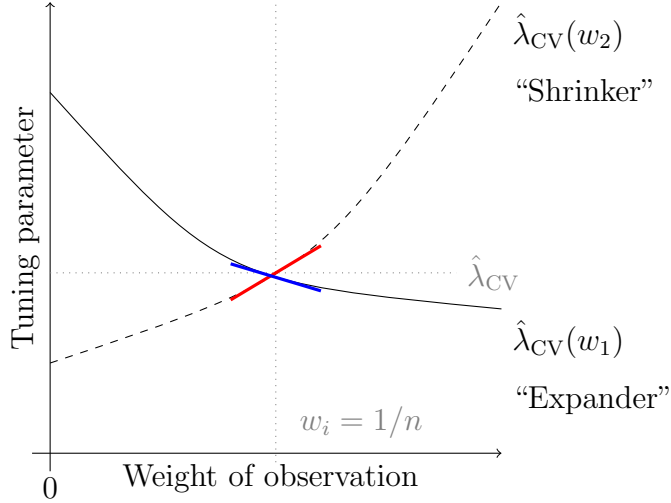
\begin{figure}%
\centering
\begin{tikzpicture}[scale=1.2]
\coordinate (origo) at (0,0);
\draw[->] (-0.2,0) -- (2.5,0) node[align=center,below] {Weight of observation } -- (5,0);
\draw[->] (0,-0.2) -- (0,0) node[below=5pt] {$0$}-- (0,2.5) node[above, rotate=90] {Tuning parameter} -- (0,5);

\draw[gray,dotted] (2.5,-0.1) node[above right = 5pt] {$ w_i = 1/n$} -- (2.5,5);

\draw (0,4) .. controls (1.9,1.9) .. (5,1.6) node[below right,align=left] {$\hat \lambda_{\cv}(w_1)$\\``Expander''};
\draw[dashed] (0,1) .. controls (3,2) .. (5,5)  node[below right,align=left] {$ \hat \lambda_{\cv}(w_2)$\\``Shrinker''};;

\draw[gray,dotted] (0,2)  -- (4.5,2) node[right] {$ \hat \lambda_{\cv}$};

\draw[very thick,red] (2,1.7) -- (3,2.3);
\draw[very thick,blue] (2,2.1) -- (3,1.8);

\end{tikzpicture}
\caption{Schematic illustration for two different observations of the change in the optimal tuning value as function of their weight in the cross-validation procedure. The continuous line shows what happens when changing the weight of observation 1 ($w_1$), the dashed one of observation 2 ($w_2$).}%
\label{fig:weightplot}%
\end{figure}

Based on the weighted cross-validation criterion, we can study how the optimal choice of the tuning parameter varies as a function of the weight of a single observation,
\begin{equation}
 \hat\lambda(w_i) = \arg\min_{\lambda>0} \wCV(\lambda,w_i). 
\label{eq:wLambda}
\end{equation}
As shown schematically in Figure \ref{fig:weightplot}, by up- and down-weighting different observations the optimal tuning value changes. For the first observation (solid line), the value of the optimal tuning parameter increases when the observation is down-weighted (until exclusion of the residual when $w_i=0$), while it decreases if the observation is copied or given more weight. Reversely, for the second observation (dashed line), the optimal tuning parameter value decreases if the observation is down-weighted (removed), while the value increases if the observation is up-weighted. 
The interpretation of these changes in the optimal value, $\hat\lambda_{\cv}$, is related to the effective degrees of freedom in the ridge model
\begin{equation*}
\df(\lambda) = \tr\left( \mat{X} ( \mat{X}^T \mat{X}+ \lambda I_p)^{-1}  \mat{X}^T \right) = \sum_{\ell=1}^p \frac{\alpha_\ell}{\alpha_\ell + \lambda},
\end{equation*}
where $\alpha_\ell$ for $\ell=1,\dots,p$ are the eigenvalues of the matrix $\mat{X}^T \mat{X}$. A larger $\lambda$ value yields fewer degrees of freedom in the model, while a smaller $\lambda$ value corresponds to more degrees of freedom. When $\lambda \to \infty$, the effective degrees of freedom approaches zero. This way the transformation $\df(\lambda)$ supplies a more intuitive and interpretable scale for the penalty parameter.

Figure \ref{fig:weightplot} demonstrates the two different types of observations: points that require a higher tuning value, i.e.\ fewer degrees of freedom when given more weight, and points that require a smaller optimal tuning value, i.e.\ more degrees of freedom in the model when given more weight. We term these two classes of points, \emph{shrinkers} and \emph{expanders}, respectively. We will characterize mathematically shrinkers and expanders in Section \ref{subsect:derivativeRidge}, but, intuitively, one can think that an observation closer to the null model (i.e., $\lambda = \infty$) than its estimated value obtained with the current model ($\lambda = \hat{\lambda}_{CV}$) is a shrinker, one farther an expander.

\subsection{Role of the derivative}\label{subsect:derivative}
As mentioned in the introduction, our approach has the advantage of describing the effect of a single observation for any modification of its contribution to the choice of the tuning parameter, as oppose to the case of its deletion. In particular, it allows us to quantify the immediate change when a point is up- or down-weighted, supplying alternative information to simply deleting observations. 

By the definition of the single weight normalized cross-validation criterion (Equation \eqref{weightedCrit}), all observations have the same weight when $w_i= \frac{1}{n}$, i.e.\ in the case of standard leave-one-out cross-validation. To analyze the immediate effect of observations up-/down-weighting, therefore, we need to study the slope of the optimal tuning parameter curve $\hat\lambda(w_i)$ (Equation \eqref{eq:wLambda}) at this specific value of the weight, i.e. 
\begin{equation*}
\left.\frac{\partial }{\partial w_i} \hat\lambda(w_i)\right|_{w_i=\frac{1}{n}}.
\end{equation*}
By implicit differentiation, this derivative with respect to the weights is, in general, proportional to the derivative of the leave-one-out error function in the value of the standard cross-validation minimum.

\begin{lemma}\label{lemma:derivative}
For a differentiable squared leave-one-out cross-validation error $e_{[i]}^2(\lambda), i=1,\dots,n$, the derivative of the optimal tuning parameter with respect to the weight of the $i$th observation is given by
\begin{equation}
\left.\frac{\partial  \hat\lambda(w_i) }{\partial w_i} \right|_{w_i= \frac{1}{n}} = -\frac{n^2 f'_{i}(\hat\lambda_{\cv} )}{(n-1)\sum_{j=1}^n  f''_{j}(\hat\lambda_{\cv})}, 
\label{eq:CVderiv}
\end{equation} 
where $f'_{i}(\hat\lambda_{\cv}) = \left. \frac{\partial e_{[i]}^2(\lambda)}{\partial \lambda}\right|_{\lambda = \hat\lambda_{\cv}}$ and $f''_{j}(\hat\lambda_{\cv}) = \left. \frac{\partial^2 e_{[j]}^2(\lambda)}{\partial \lambda^2}\right|_{\lambda = \hat\lambda_{\cv}}$. 
\end{lemma}
The proof of Lemma \ref{lemma:derivative} can be found in Section S.2 in the Supplementary material. Let us focus here on its interpretation. Equation \eqref{eq:CVderiv} tells us that the influence of a single observation is determined by the derivative of its penalized error function computed at the value determined by all observations ($\hat\lambda_{\cv}$). The contribution of the second derivative, on the other hand, does not affect the relative influence of the observations, as the denominator is equal for all observations. 

\subsection{Derivative in the ridge regression case}\label{subsect:derivativeRidge}
For ridge regression, by Equation \eqref{eq:CVderiv}, the derivative of $\hat\lambda(w_i)$ is
\begin{align*}
\hspace{-1cm} &\left.\frac{\partial  \hat\lambda(w_i) }{\partial w_i} \right|_{w_i= \frac{1}{n}} \propto \frac{y_i- \vc{x}_{i}^T(\mat{X}^T\mat{X} + \hat\lambda_{\cv} I_p)^{-1} \mat{X}^T\vc{y}}{(1-\vc{x}_{i}^T(\mat{X}^T \mat{X} + \hat\lambda_{\cv} I_{p})^{-1}\vc{x}_{i})^3}
\bigg[ \vc{x}_{i}^T(\mat{X}^T\mat{X} + \hat\lambda_{\cv} I_p)^{-2} \mat{X}^T\vc{y}(1 - \\
& \quad \vc{x}_{i}^T(\mat{X}^T \mat{X} +  \hat\lambda_{\cv} I_{p})^{-1}\vc{x}_{i}) - (y_i- \vc{x}_{i}^T(\mat{X}^T\mat{X} +  \hat\lambda_{\cv} I_p)^{-1} \mat{X}^T\vc{y}) \vc{x}_{i}^T(\mat{X}^T \mat{X}+ \hat\lambda_{\cv} I_{p})^{-2}\vc{x}_{i} \bigg].
\end{align*}
Let us assume a single standardized covariate, $\sum_{i=1}^n x_{i}^{ 2} = 1, \bar x = 0$, and $\bar y = 0$. While this situation is not of practical interest, it is useful to illustrate the role of the derivative in a simplified setting. By denoting the OLS residual $r_i = y_i -\hat y_i(0)$ and the OLS leverage $h_{ii}= H_{ii}(0)$, the leave-one-out error is given by
\begin{equation*}
f_{i}(\lambda) = \left( \frac{ \lambda y_i + r_{i}}{\lambda + 1 - h_{ii}} \right)^2,
\end{equation*}
and we have the first derivative with respect to $\lambda$,
$
f_{i}'(\lambda) = \frac{-2 (\lambda y_i+ r_i) (r_i - y_i(1-h_{ii}))}{(\lambda + 1 - h_{ii})^3}.
$
Together with Equation \eqref{eq:wLambda} this gives
\begin{equation}
\left.\frac{\partial }{\partial w} \hat\lambda(w)\right|_{w=\frac{1}{n}} = \frac{2n^2 (r_i + \hat\lambda_{\cv} y_i) (r_i-y_i (1-h_{ii}) )}{(n-1)( \hat\lambda_{\cv} + 1- h_{ii})^3 \sum_{j=1}^n  f''_{j}(\hat\lambda_{\cv})}.
\label{eq:CVderiv2}
\end{equation} 
Based on Equation \eqref{eq:CVderiv2} we can say something regarding the relative influence of each observation on the choice of the tuning parameter, and, importantly, identify  whether an observation is a shrinker or an expander by evaluating its sign. As seen in Figure \ref{fig:weightplot}, a negative derivative indicates a shrinker, while a positive derivative indicates an expander.

\begin{figure}%
\centerline{
\begin{tikzpicture}[scale=0.8]
\draw[->] (-3,0) -- (3,0) node[align=center,right] {$x$} ;
\draw[white, fill=gray,pattern=north east lines, pattern color=gray]  (0,0) -- (3,2) --(0,2) -- (0,0) -- cycle;
\draw[white,fill=gray,pattern=north east lines, pattern color=gray]  (0,0) -- (-3,-2) --(0,-2) -- (0,0) -- cycle;
\draw[dashed] (-3.2,-2.2) -- (3.2,2.2) node[below=10pt] {$y=\frac{\hat{\beta}}{(1+\hat\lambda_{\cv})}x$};
\draw[->] (0,-2.1) -- (0,2.1)  node[above] {$y$};
\end{tikzpicture}
\begin{tikzpicture}[scale=0.8]
\draw[->] (-3,0) -- (3,0) node[align=center,right] {$x$} ;
\draw[white,fill=gray,pattern=north east lines, pattern color=gray]  (0,0) -- (3,-2) --(0,-2) -- (0,0) -- cycle;
\draw[white,fill=gray,pattern=north east lines, pattern color=gray]  (0,0) -- (-3,2) --(0,2) -- (0,0) -- cycle;
\draw[dashed] (-3.2,2.2) -- (3.2,-2.2) node[above=12pt] {$y=\frac{\hat{\beta}}{(1+\hat\lambda_{\cv})}x$};
\draw[->] (0,-2.1) -- (0,2.1)  node[above] {$y$};
\end{tikzpicture}}
\caption{Illustration of areas where the expanders can be found (shaded). The left panel shows a ridge regression line with positive slope, while the right panel shows a regression line with negative slope.}%
\label{fig:DestabPlot}%
\end{figure}
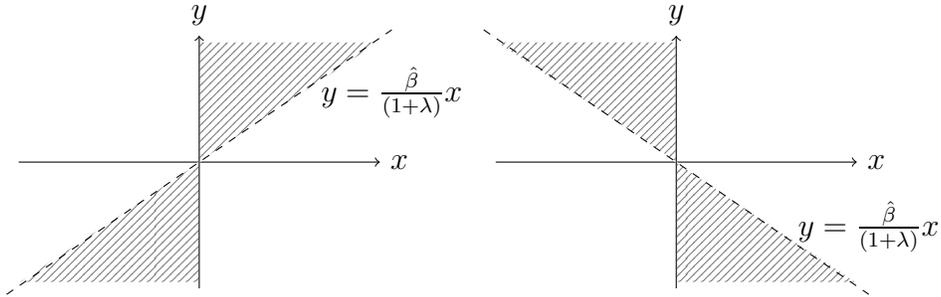

As the leverage assumes values between 0 and 1 ($0<h_{ii}<1$), and the tuning parameter is positive ($\hat\lambda_{\cv}\geq 0$), the factor $(\lambda + 1- h_{ii})^3 $ in the denominator of Equation \eqref{eq:CVderiv2} is also positive. Further, the sum of second derivatives of the cross-validation error, $\sum_{j=1}^n f''_{j}(\hat\lambda_{\cv})$, must be positive because $\hat\lambda_{\cv}$ determines a minimum. The sign of Equation \eqref{eq:CVderiv2} is thus determined by the nominator alone,
\begin{equation}
\label{eq:destabilizingfactor}
    (r_i + \hat\lambda_{\cv} y_i) (r_i-y_i (1-h_{ii}) ).
\end{equation}
In order for an observation to be an expander, the expression must be negative and its two factors must have opposite signs. This is achieved when the outcome $y_i$ and the residual $r_i$ satisfies
\begin{equation}
\label{eq:destabilizer1dim}
    -\hat\lambda_{\cv} < \frac{r_i}{y_i} < 1-h_{ii}.
\end{equation}
The derivation of this result can be found in Section S.3 in the Supplementary material. Based on the inequalities in \eqref{eq:destabilizer1dim}, we can give intuitive conditions for when an observation is an expander.  As $r_i=y_i-x_i\hat{\beta}=y_i - x_i(\sum_{j=1}^{n}x_jy_j)= y_i - x_i(x_iy_i+\hat{\beta}_{[i]})$, in the case that $y_i>0$, the right-side inequality in \eqref{eq:destabilizer1dim} becomes $x_i \hat{\beta}_{[i]} > 0.$ If $\hat{\beta}_{[i]}>0$, we must then have that $x_i>0$, implying that the observation must be in the first quadrant, and $\hat{\beta} = \hat{\beta}_{[i]}+x_iy_i>0$, such that the OLS coefficient must be positive as well. Moreover, the left inequality in \eqref{eq:destabilizer1dim} becomes
\begin{equation}\label{eq:destabilizerRegIneq}
    y_i > \frac{x_i\hat{\beta}}{1 + \hat\lambda_{\cv}},
\end{equation}
which is equivalent to having a positive ridge residual for the observation $i$. In summary, an observation with a positive $y$ value can only be an expander if $x$ is also positive, i.e.\ the observation is in the first quadrant, and its outcome lies above the ridge regression line, as it can be seen in the left panel of Figure \ref{fig:DestabPlot}. If, instead, $\hat{\beta}_{[i]}<0$, we have $x_i<0$ and the observation must be in the second quadrant. Then $\hat{\beta}=\hat{\beta}_{[i]}+x_i y_i<0$, and we get that the observation must satisfy \eqref{eq:destabilizerRegIneq} as seen in the right panel of Figure \ref{fig:DestabPlot}. The observation with a positive $y_i$ and negative $x_i$ is an expander for a positive ridge residual. 

For the opposite case, where $y_i<0$, the inequalities in (\ref{eq:destabilizer1dim}) change to $x_i \hat{\beta}_{[i]} < 0$ and $y_i < \frac{x_i\hat{\beta}}{1+\hat\lambda_{\cv}}$, and the picture reverses. For $\hat{\beta}>0$, the observation, $x_i<0$ lies in the third quadrant and the residual is negative (left panel of Figure \ref{fig:DestabPlot}). While for $\hat{\beta}<0$, the observation $x_i>0$ is in the fourth quadrant also with a negative ridge residual (right panel of Figure \ref{fig:DestabPlot}). In Figure \ref{fig:DestabPlot}, all the areas where observations will be defined as expanders (regardless of their effective influence) are shaded. 

\subsection{Graphical investigation}\label{subsect:graphical}
To better grasp the influence of the single observations on the choice of the tuning parameter, we propose to plot the curves of the optimal tuning parameter $\hat{\lambda}(w_i)$ (Equation \eqref{eq:wLambda}) as a function of the weight for each observation (see, e.g., Figure \ref{fig:Simulations}). The horizontal axis gives the weight of the observations relative to $1/n$, where all curves meet in the factor of $1$ marked by the vertical line. This is the point where we chose to analyze the slope of the curves (see Sections \ref{subsect:derivative} and \ref{subsect:derivativeRidge}), and such a plot is an effective tool to visualize the observations' impact. Steeper curves mean higher impact; as a shrinker if the function is increasing, and as an expander in the opposite case. This follows in the tradition of \citet{genton2010visualizing}, where an observation is influential ``whenever a change in its value leads to a radical change in the estimate'' and the graphical hair-plot is used for visual identification.

The observations' influence on the choice of the tuning parameter can be evaluated globally, and a curve strongly separated from the others is an indication of the presence of an influential point. Note that the left boundary, $w_i = 0$ represents the complete exclusion of the residual, while the right margin is arbitrary: in our examples we have chosen to stop at 4 (which means to replicate the observation approximately 4 times in the sample), but nothing prevents from extending the limit (see the lower panel of Figure S.4 in the Supplementary material for an example). This visual tool to asses the influence of the observations is implemented in the R package \emph{influridge}, available on GitHub (\url{www.github.com/khellton/influridge}). 

\section{Simulated data examples}\label{sect:simulations}
We now use simulated data to illustrate our approach. In order to have realistic high-dimensional data, we use as input the gene expression profiles ($p=19411$) of $40$ samples of glioma progenitor cells\footnote{The data are available in the EMBL-EBI ArrayExpress database (\texttt{www.ebi.ac.uk/arrayexpress}) under accession number E-GEOD-76990.}. \citet{moeckel2014response} collected them to study whether genes held predictive information regarding the outcome for tyrosine kinase inhibitor (Sunitinib) treatment of glioblastoma cancer cells. In our study, we instead create simulated outcomes $y_i = \boldsymbol{x}_i^T\boldsymbol{\beta}+\varepsilon_i,$ where $\varepsilon_i\sim\mathcal{N}(0,1)$ for  $i=1,...,n$. For easy visualization of the high-dimensional data, the regression coefficients $\boldsymbol{\beta}$ are constructed to be a small perturbation of the first principal component loadings of the genomic data. In addition to the curves of $\hat{\lambda}(w_i)$, we report a scatter plot of the outcome against the first principal component, which explains $24.72\%$ of the total variance of the original data. The scatter plots also show the regression line (dashed) of the outcome against the first principal component and the horizontal line (dotted) of the mean outcome. The curves and corresponding points of notable observations are highlighted.

\paragraph{Shrinkers vs expanders.}

\begin{figure}%
\centering
  \includegraphics[width=0.49\textwidth]{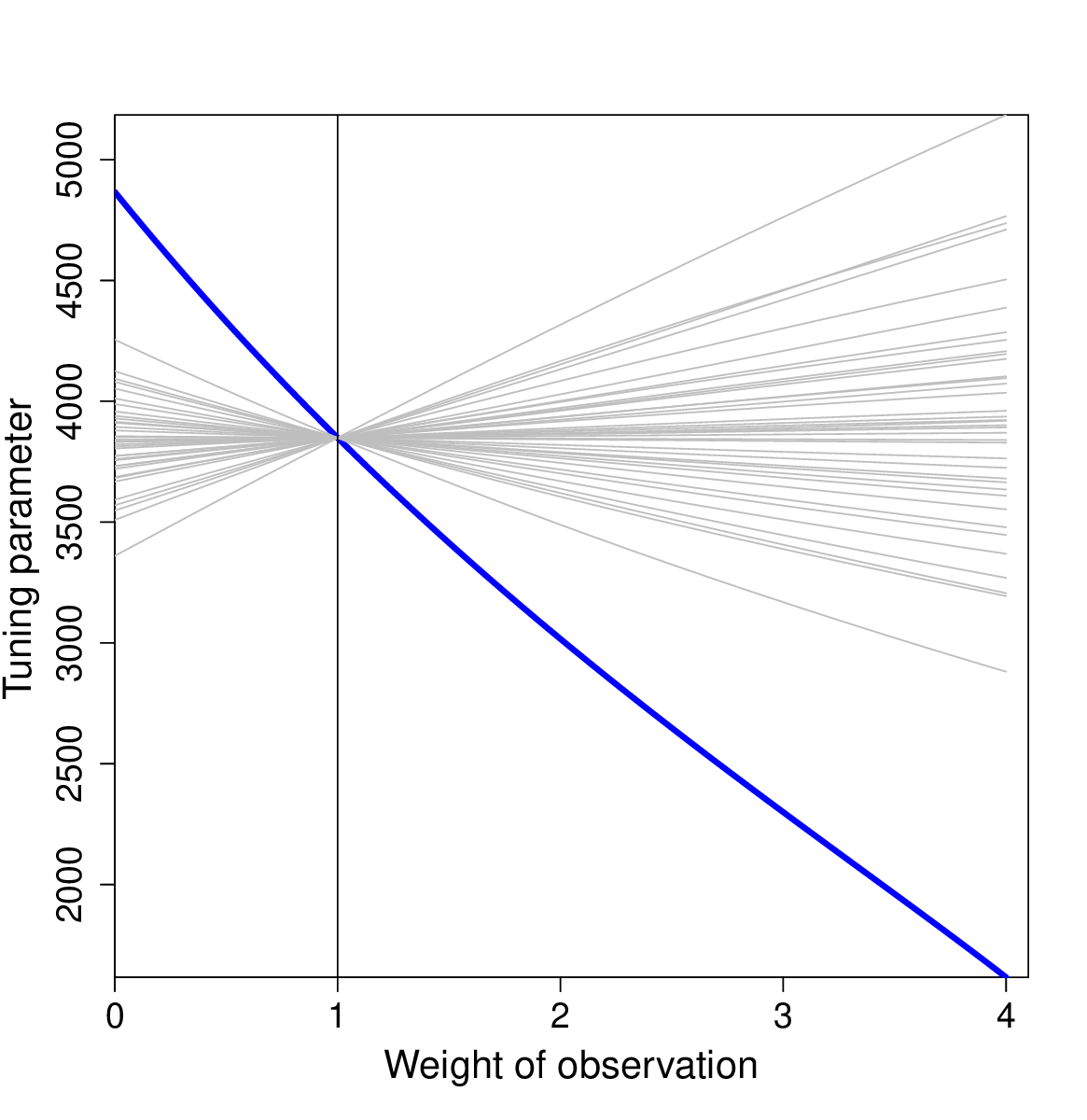}
  \includegraphics[width=0.49\textwidth]{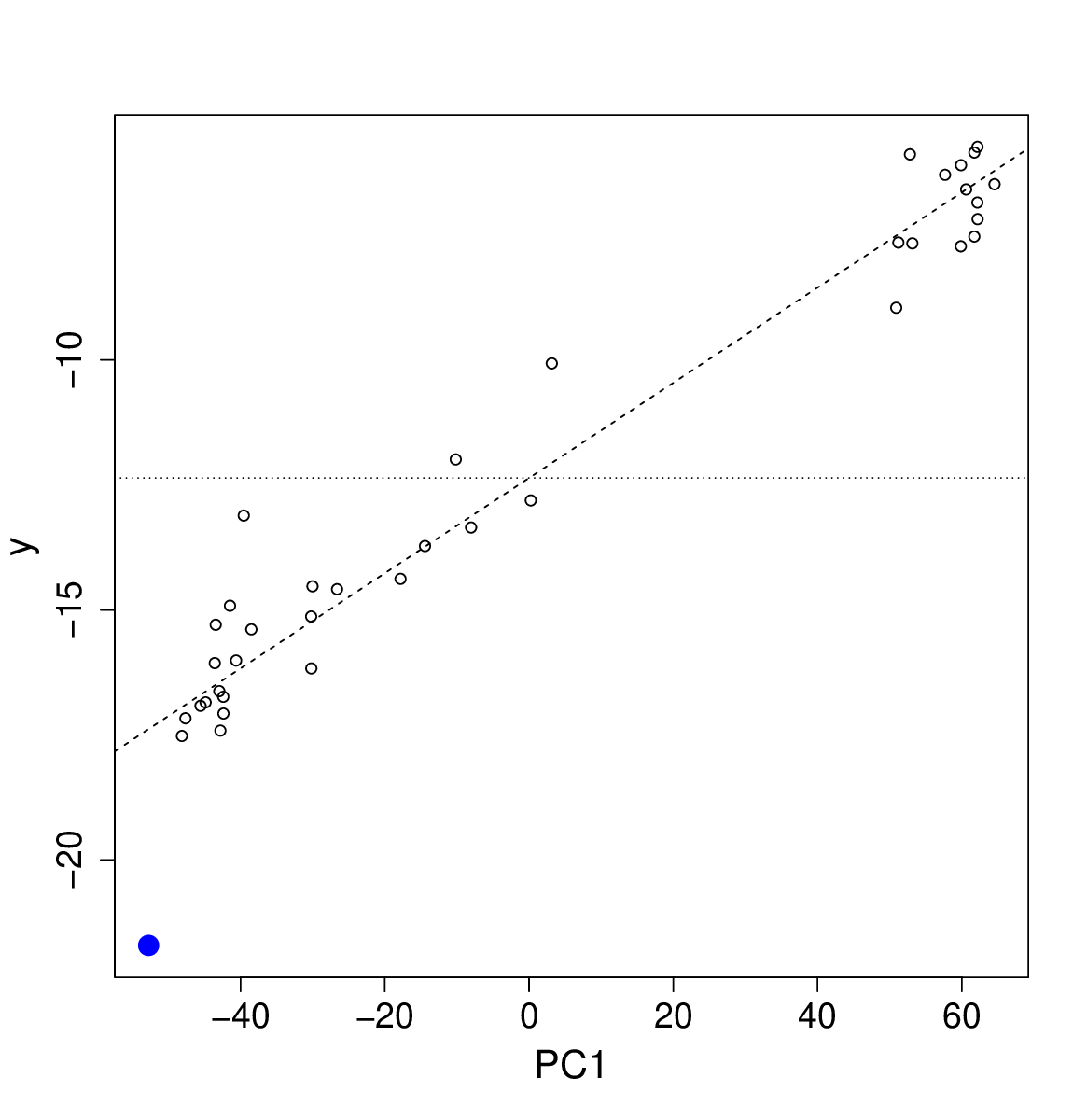}
  \includegraphics[width=0.49\textwidth]{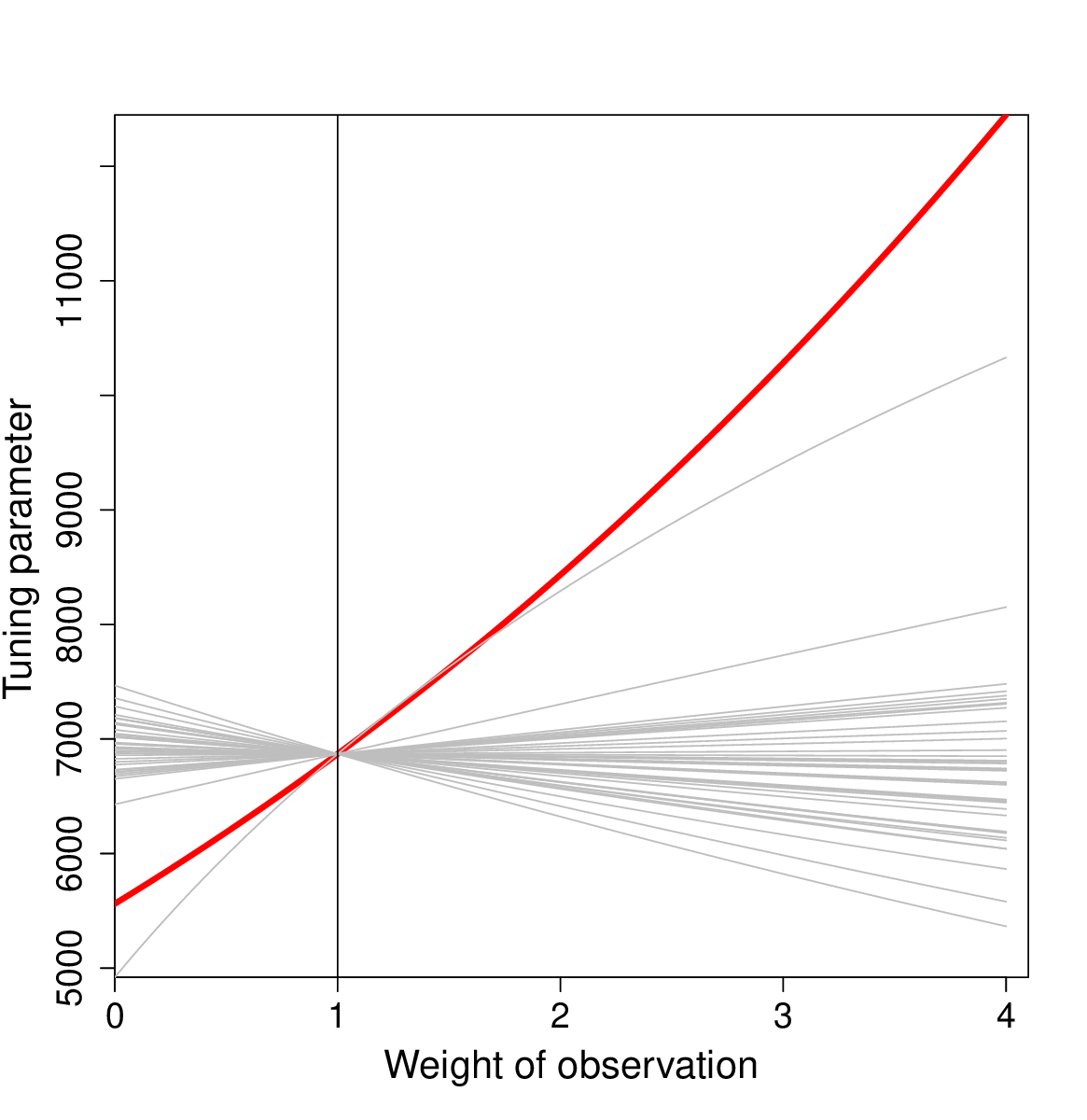}
  \includegraphics[width=0.49\textwidth]{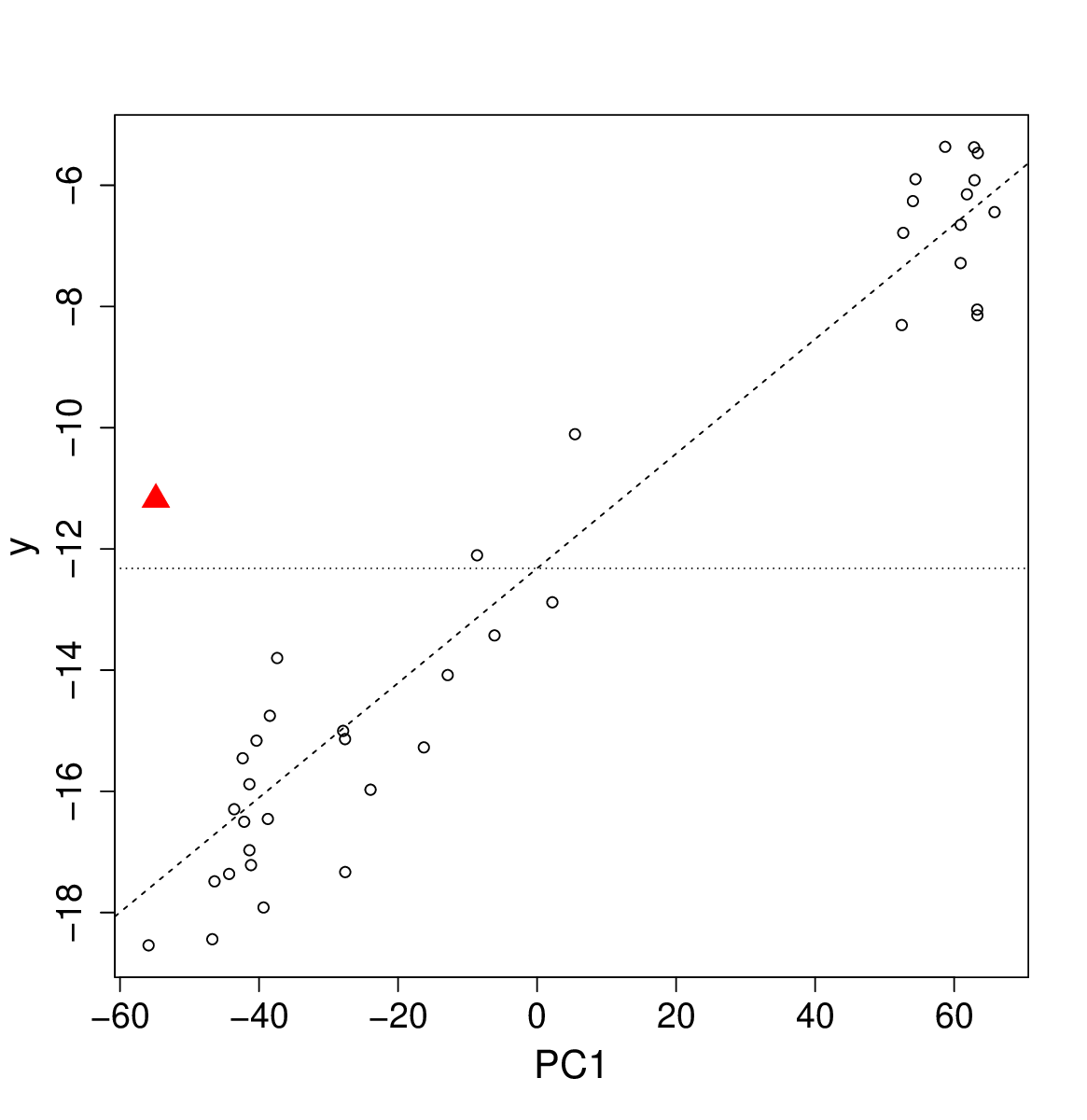}
\caption{Curves of $\hat\lambda_{\cv}(w_i)$ as functions of $w_i$ for the 40 observations of the simulated data examples (left panels) and scatter plots of their responses against the first PC (right panels). Top row: presence of an expander (bold circle); bottom row: presence of a shrinker (bold triangle).}
\label{fig:Simulations}
\end{figure}

Figure \ref{fig:Simulations} shows the typical behavior of the $\hat{\lambda}(w_i)$ curves in the presence of an expander (top row) and a shrinker (bottom row), artificially created by adding to the observation a large negative (expander) or positive (shrinker) residual. The $\hat{\lambda}(w_i)$ curve of the modified observation (bold line) is significantly steeper than the other curves, making it easy to identify it as an influential point. Although this is not a one-dimensional setting, the PCA plots (right panels of Figure \ref{fig:Simulations}) resemble the situation in the left panel in Figure \ref{fig:DestabPlot}, with the points under investigation having a negative covariate value and an outcome below (expander) or above (shrinker) the (increasing) regression line. It also helps to visualize the intuitive explanation described at the end of Section \ref{subsect:shrinkersExpanders}.

The bottom panel shows the effect of the presence of a shrinker. This particular simulation case also illustrates an important advantage of our method with respect to traditional, dichotomous exclusion/inclusion based approaches. Implementing the latter strategy, indeed, the observation we constructed to be a shrinker would not be considered the most influential. When the weight is equal to 0, in fact, there is another observation pointing to the smallest value of the penalty term (bottom left plot, Figure \ref{fig:Simulations}). Only taking the whole curve into consideration (and in particular its steepness) one can correctly identify it. There is, of course, no 'correct' answer to how an observation is influential, but it is worth noting that our approach provides information unavailable to the traditional deletion approaches.

The Supplementary material (Figure S.1) contains three other examples; 1) no influential points, 2) both a shrinker and an expander, and 3) a combination of shrinkers and expanders, are present in the data. The latter two cases, in particular, show how an observation can be a shrinker simply by opposing the effect of the expander, illustrating the importance of interaction between observations to determine their relative influence. 

\paragraph{Comparison with \cite{RajaratnamAl2019}.}\label{subsec:simulComp}
As a comparison, we applied the influence-lasso method of \cite{RajaratnamAl2019}, briefly described in the introduction, to our simulated examples. The results are reported in the Supplementary material (Table S.1) and show that it is not straightforward to only identify as influential points the observations that were in fact simulated to be influential.  Influence-lasso, in particular, tends to select more observations as influential than those that truly are. In contrast to our visual approach, influence-lasso has a test-based procedure to select the observations, and some regular points are expected to be incorrectly marked as influential due to the type-I error. Paradigmatic is the example with no simulated influential points (Table S.1, first row): among the 40 simulated (under the null hypothesis) observations, there are exactly 2 (i.e., 5\%) false positives. 

\begin{figure}%
\centering
  \includegraphics[width=0.41\textwidth]{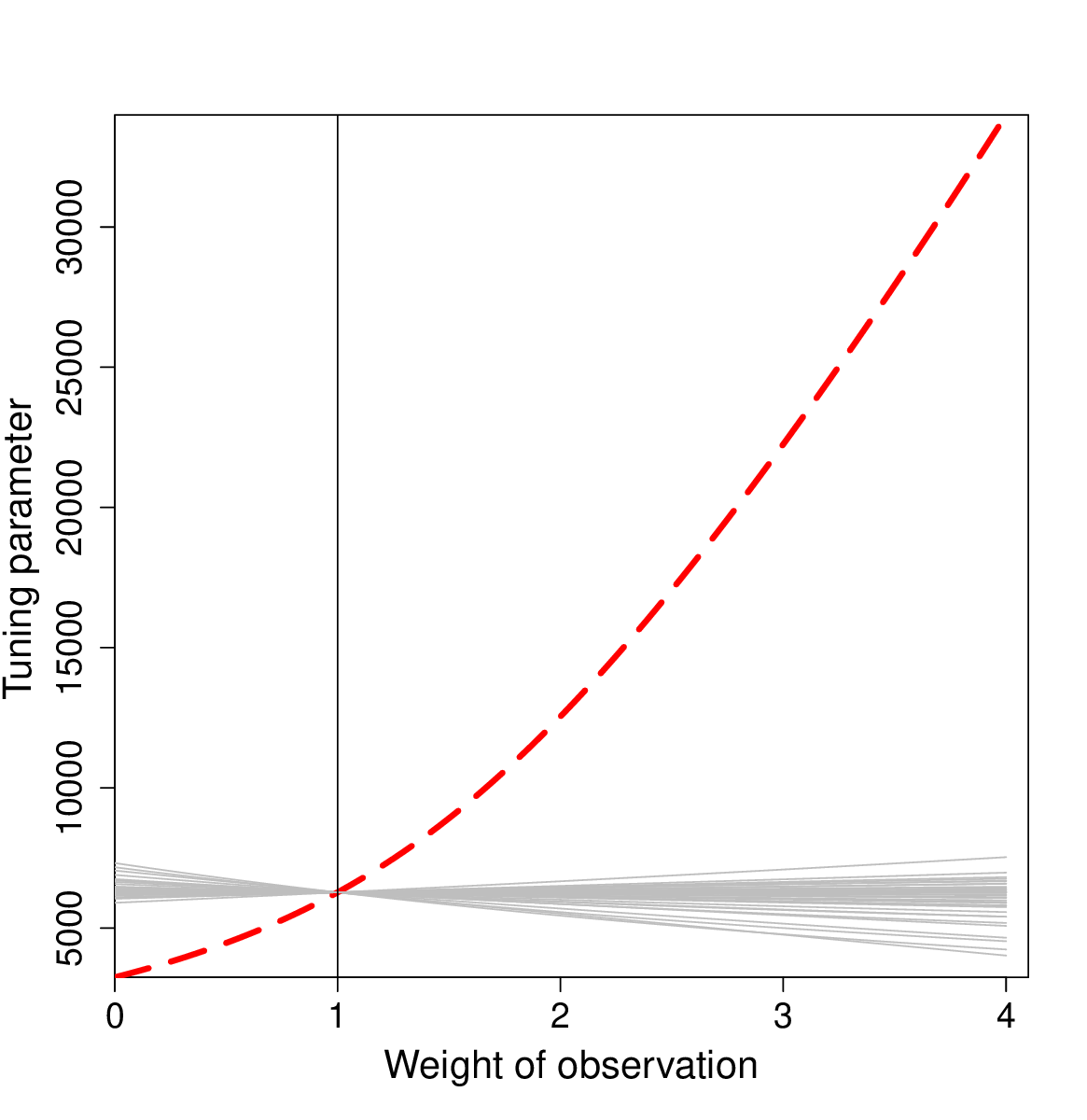}
  \includegraphics[width=0.41\textwidth]{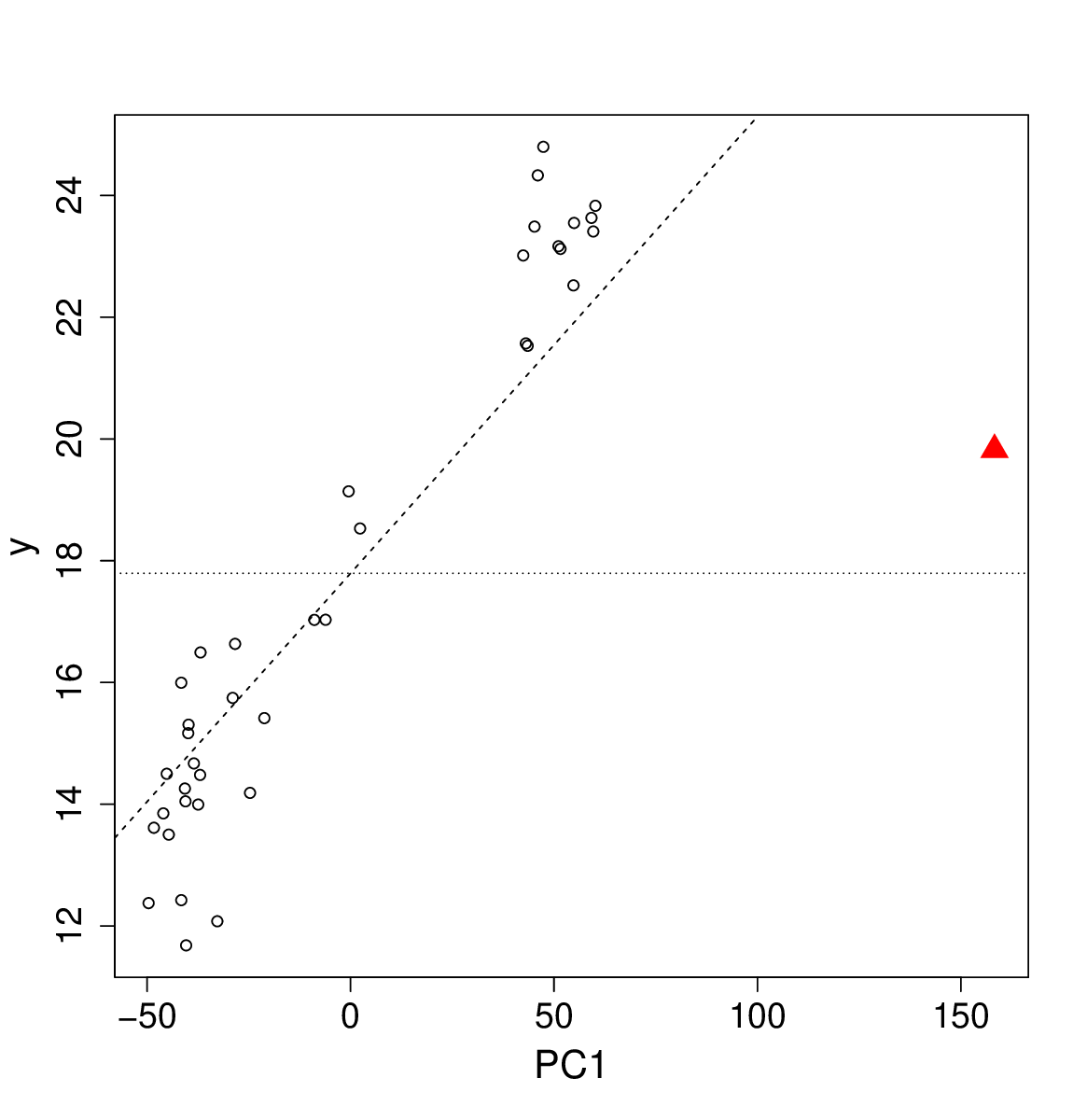}
  \includegraphics[width=0.41\textwidth]{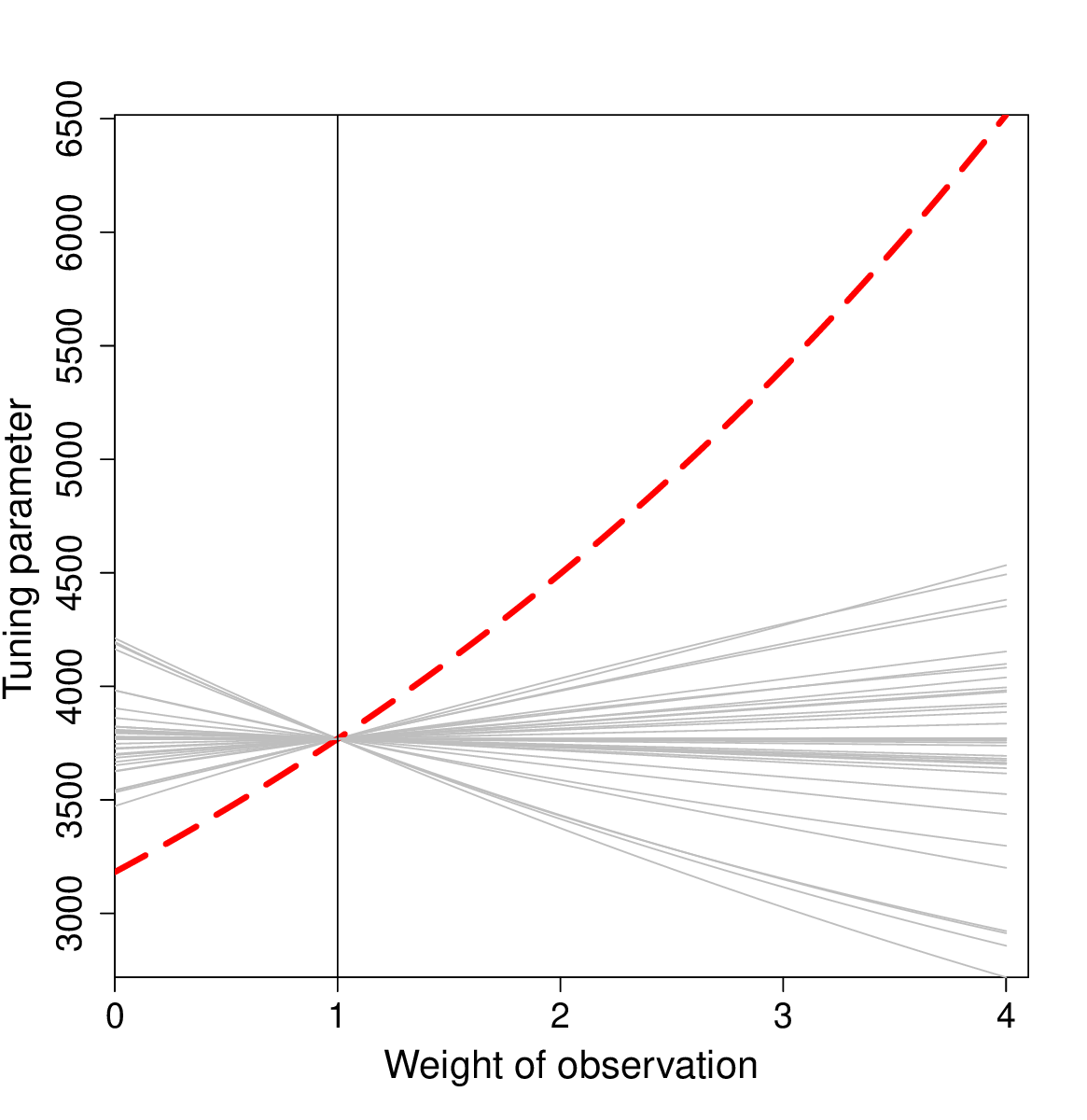}
  \includegraphics[width=0.41\textwidth]{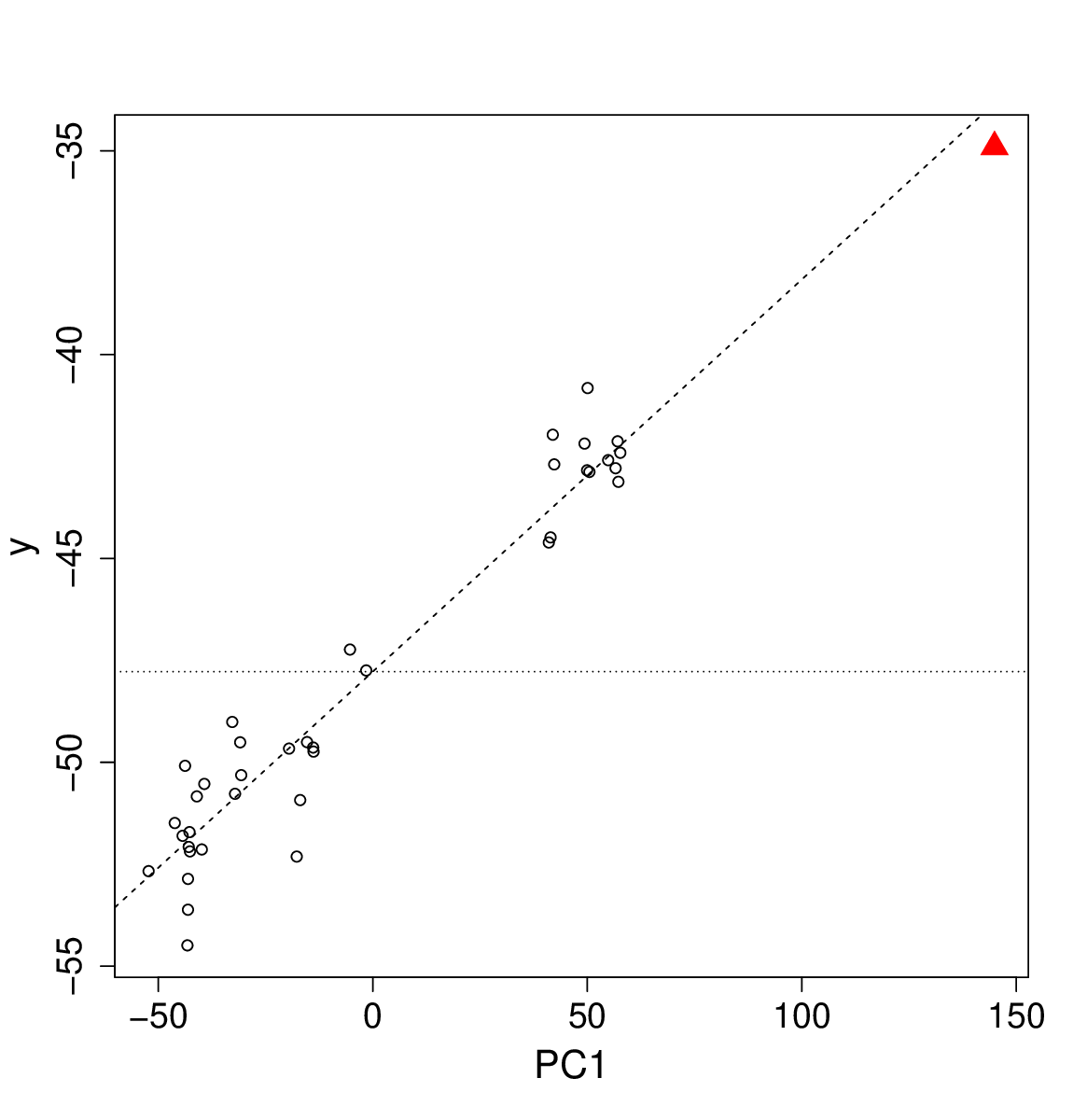}
  \includegraphics[width=0.41\textwidth]{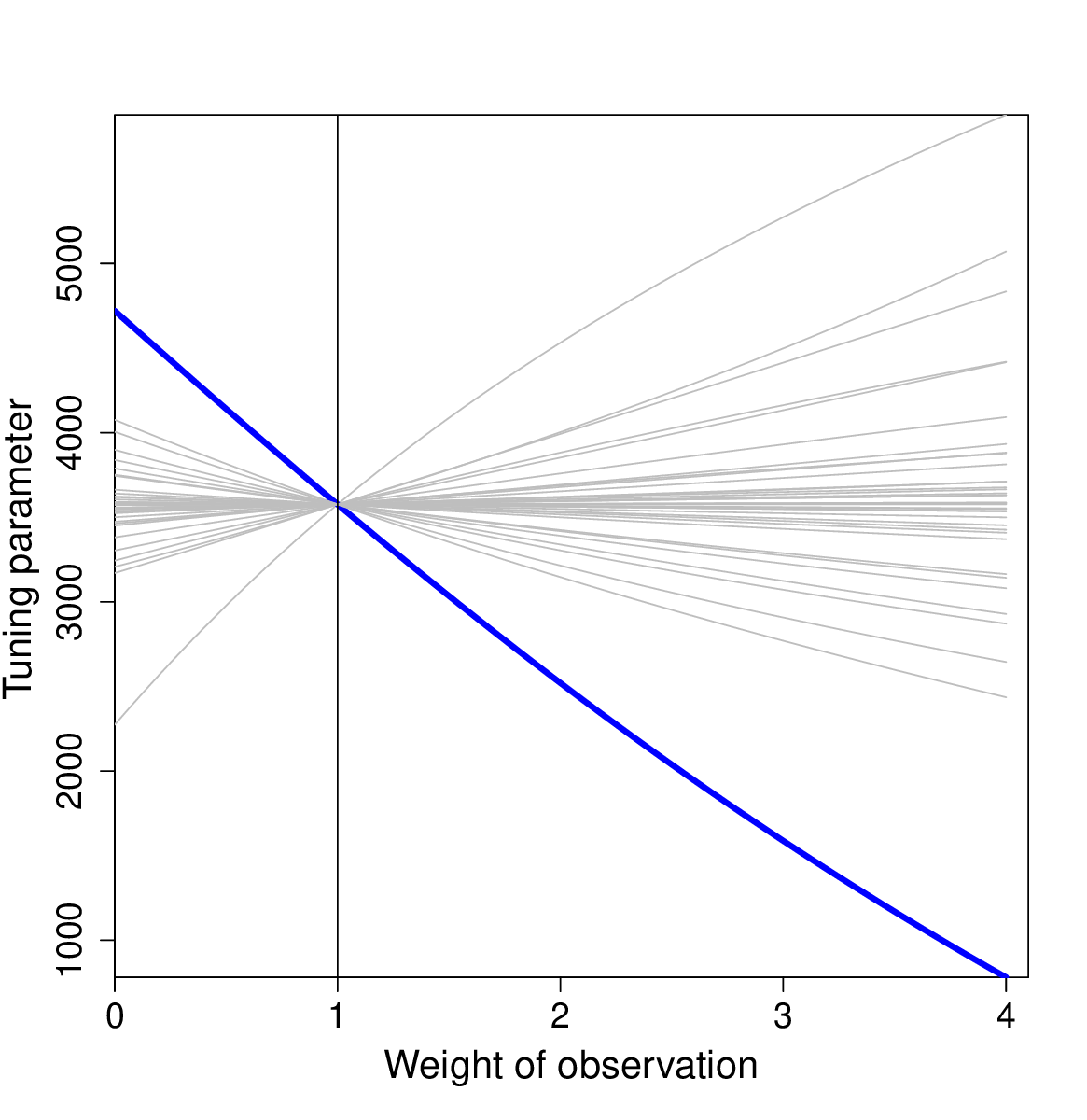}
  \includegraphics[width=0.41\textwidth]{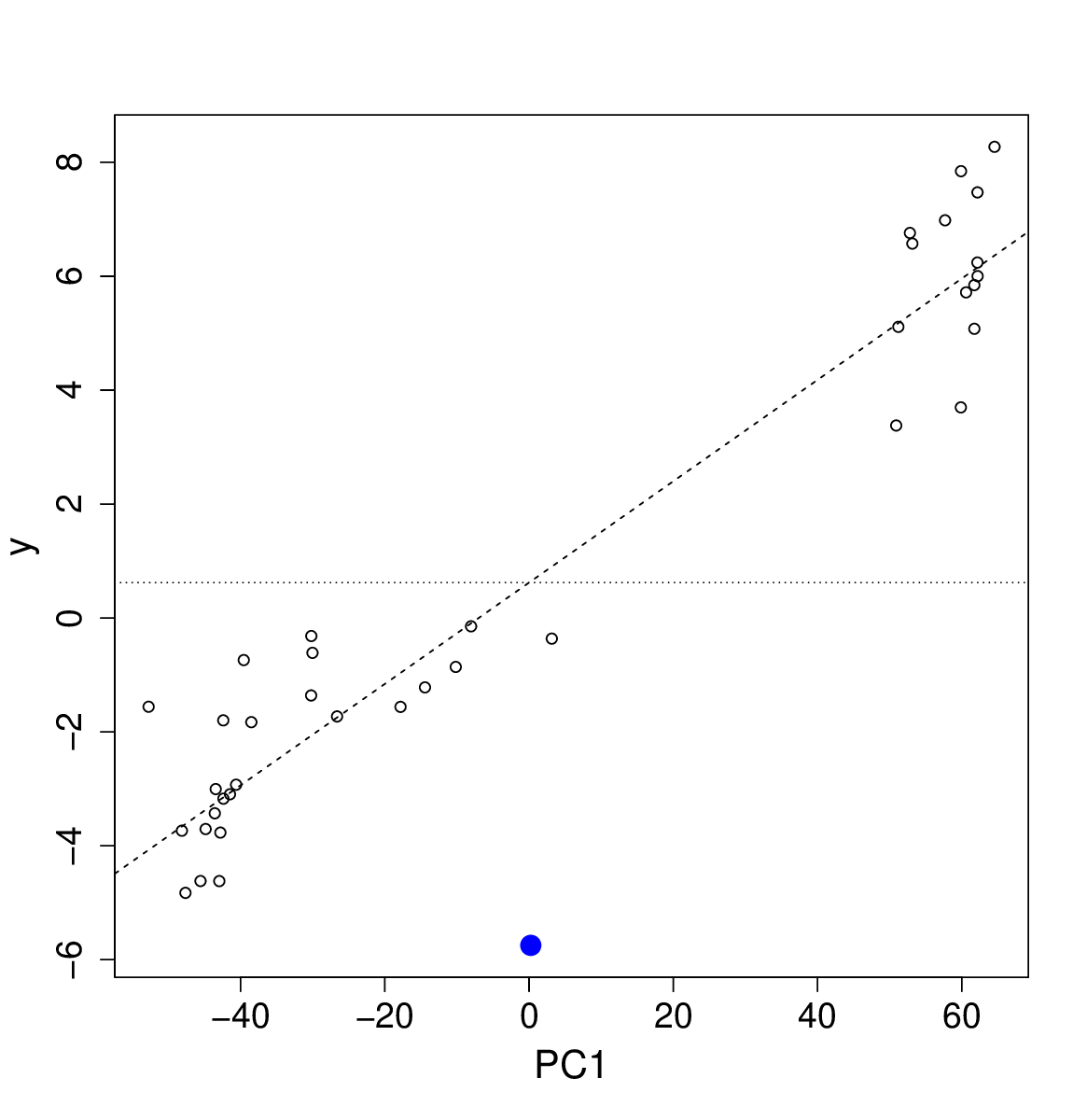}
\caption{\label{fig:leverage}: Examples of influence of some points based on their residuals and leverages: (top panels) large leverage, large residual; (middle panels) large leverage, small residual; (bottom panels) small leverage, large residual.}
\end{figure}

\paragraph{Residuals and leverage.} 
The computations of Section \ref{subsect:derivativeRidge} indicate that the influence of the single observations depends on leverages and residuals. Figure \ref{fig:leverage} shows that large values for leverage and residual generally lead to a strong influential point (top row), and that a large value of only one of them may be sufficient to characterize and observation as influential (second and third row). As we can notice by looking at the y-axis, however, the influence is smaller. When the residual or the leverage are too small, the observation may not be influential at all, no matter of the value of the other quantity (see, as an example, Figure S.2 in the Supplementary material).

\paragraph{Increasing number of influential points.} 
Finally, Figure S.3 in the Supplementary material shows the behavior of our approach with an increasing number of influential points. When one or two points have the same influence (top row) on the choice of the tuning parameter, no matter if they are shrinkers or, as in the figure, expanders, the corresponding $\hat{\lambda}(w_i)$ curves clearly stand out from the other. When their number increases (Figure S.3, middle and bottom rows), their nature of shrinkers or expanders is maintained, as they keep ``asking'' for more or less penalty when down-weighted. Their singular effect exceeding that of the similar observations, instead, gets smaller and smaller, until they cannot considered influential points anymore.

\section{Real data examples}\label{sect:realData}

\subsection{Low dimensional case: educational body fat}
\label{sec:bodyfat}
A patient's general health status can be assessed by considering a measure of body fat \citep{myint2014body}. \citet{Johnson96Bodyfat} measured the percentage of body fat of $252$ men by an underwater weighing technique, along with their age, weight, height and ten continuous body circumference measurements. The dataset is publicly available\footnote{\texttt{http://lib.stat.cmu.edu/datasets/bodyfat}} and it is known to contain one strong influential point, observation $39$. For example, while modeling the relationship between outcome and variables by fractional polynomial functions, \citet{royston2007improving} found that observation $39$ highly influences the choice of the function. \citet{DeBin17Influential} re-analyzed the data assuming linear effects, and also found that the $39$th observation was highly influential in the choice of the model. In addition, they identified observation $221$ as influential, with the opposite effect of observation $39$ on the results of the variable selection procedure.

We wish to investigate whether observations $39$ and $221$ are also influential in the choice of the tuning parameter. Although the application is low-dimensional ($p<n$), it provides insight when compared to previous studies. As the age covariate was measured on a categorical scale, it was 
omitted for simplicity from the analysis. Further, all variables were standardized to have unit variance. 

\begin{figure}%
\centering
  \includegraphics[width=0.49\textwidth]{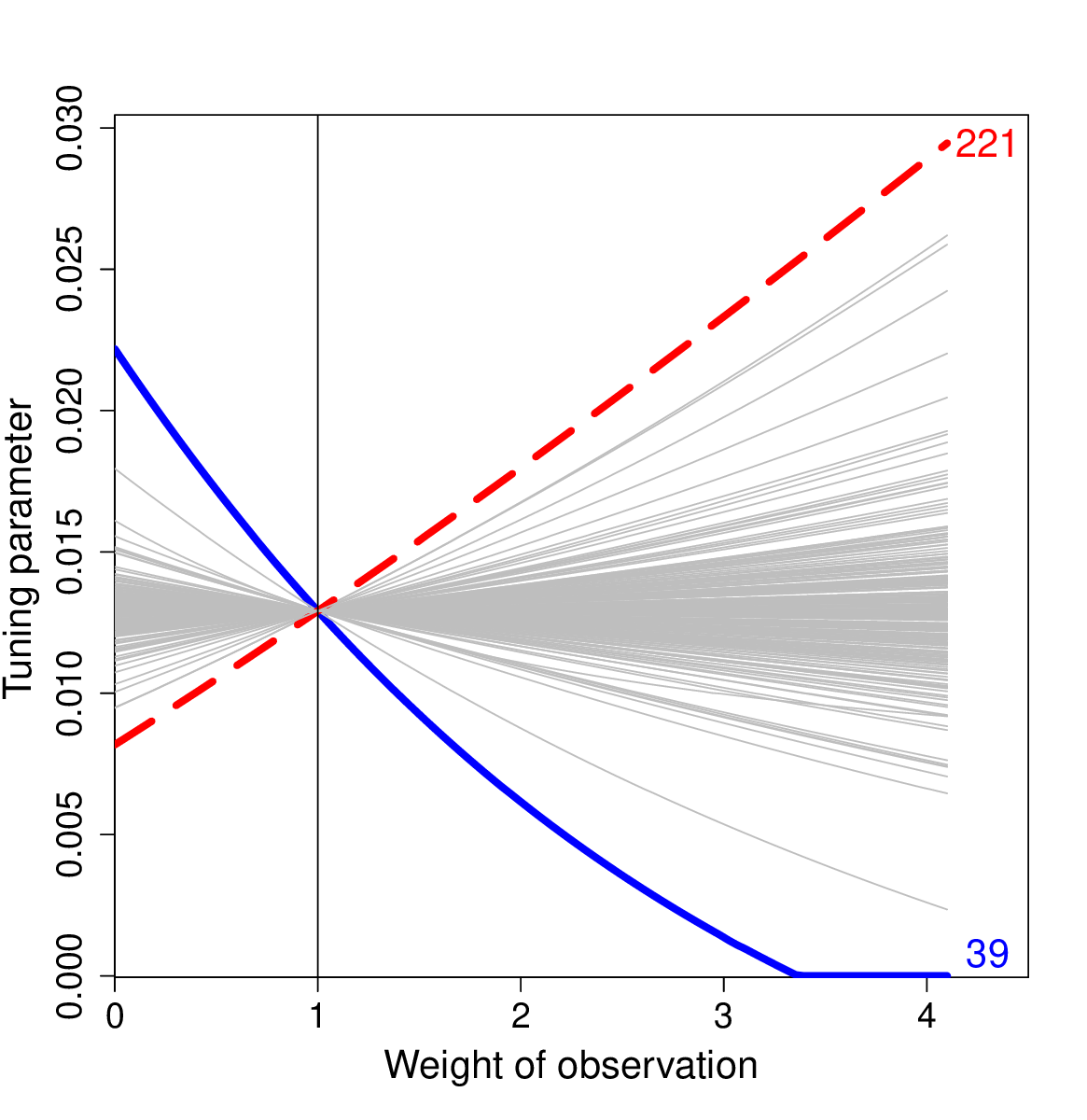}
  \includegraphics[width=0.49\textwidth]{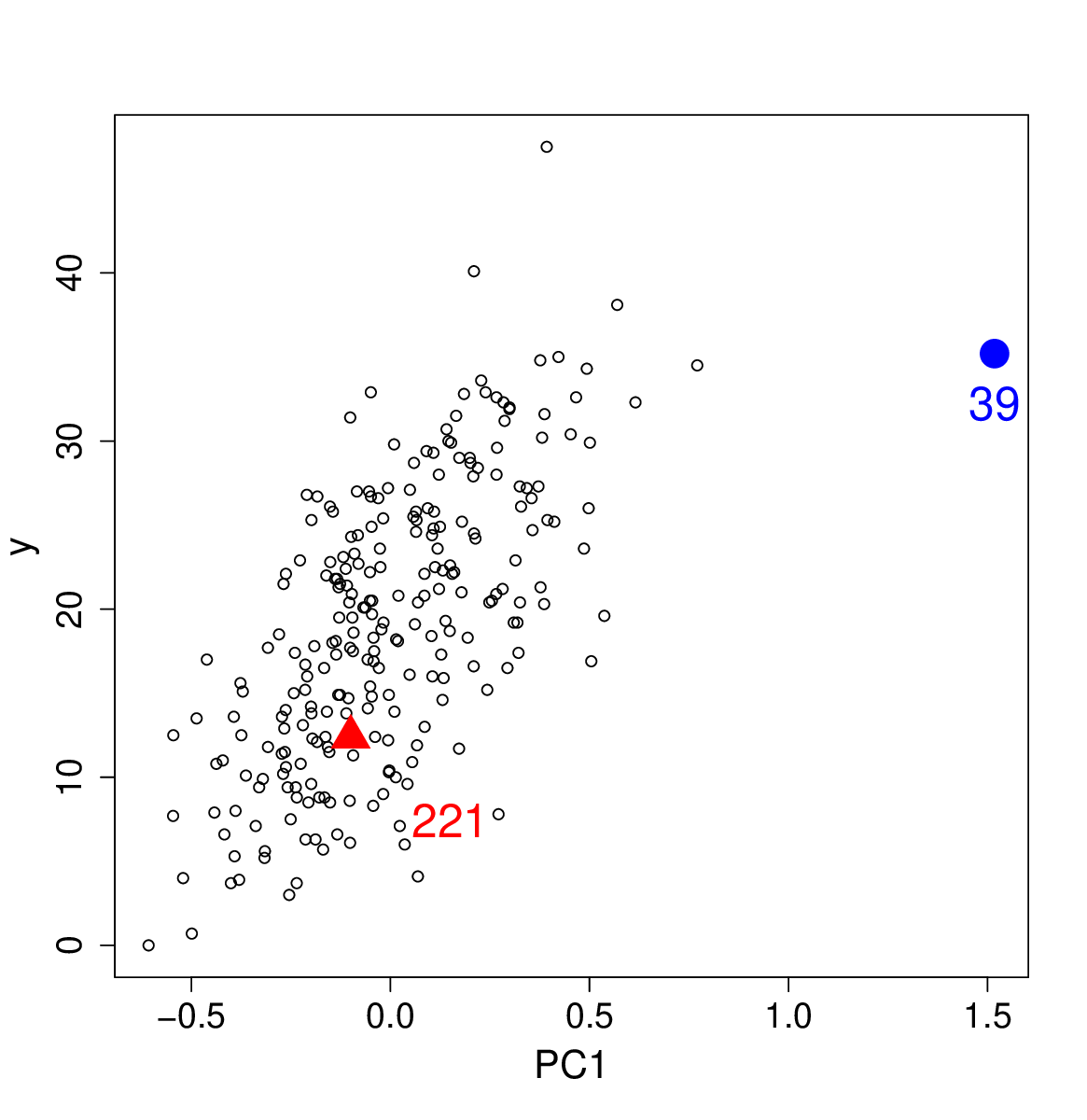}
\caption{Curves of $\hat\lambda_{\cv}(w)$ as functions of $w_i$ for the observations in the body fat dataset and scatter plot of their responses against the first principal component.}
\label{fig:BodyfatScatter}
\end{figure}

For each observation, we calculated the optimal cross-validation tuning parameter value, $\hat\lambda_{\cv}(w_i)$, as a function of the weight assigned to the observation. The resulting curves are shown in Figure \ref{fig:BodyfatScatter}. The curves of observations $39$ and $221$ are reported in bold and they are both visibly the steepest (though in different directions), meaning they change $\hat\lambda_{\cv}$ more than the other observations when perturbing their weights. The two observations are thus found to be the most influential with regard to effecting model complexity, agreeing with the results of \citet{DeBin17Influential}.

Figure \ref{fig:BodyfatScatter} includes a scatter plot of the outcome against the first principal component, which explains $77.62\%$ of the total variance. Notably, the covariates of observation $39$ stand out and strongly deviate from the rest of the observations, which in part explains its influence. On the other hand, observation 221 is very close to the average observation. Note that when observation 39 is given more weight, $\hat\lambda_{\cv}$ decreases as the model complexity required to sufficiently explain the data increases. At a certain point, around weights of $3/n$ and $4/n$, approximately corresponding to a tripling or quadrupling the observation in the data, it forces the model to be as large as possible, meaning $\hat\lambda_{\cv}=0$ or unpenalized OLS. In contrast, when observation $221$ is given more weight, $\hat\lambda_{\cv}$ increases as the model complexity required to sufficiently explain the data decreases. Observation $39$ is here an expander, while observation $221$ works as a shrinker. 

Note that the curve of observation $39$ is steeper than that of observation $221$, highlighting the larger impact of the former with respect to the latter. This is in line with the findings of \citet{DeBin17Influential}. Moreover, Figure \ref{fig:BodyfatScatter} also explains both why \citet{DeBin17Influential} identified that observation $221$ affects the inclusion of some variables in opposite ways of observation $39$ (their $\hat\lambda_{\cv}(w_i)$ curves move in opposite directions) and possibly why it was not identified by \citet{royston2007improving}. Based on the first principal component, observation $221$ is close to many other points, therefore it has no influence on the functional form. While \citet{DeBin17Influential} simply concluded that observation $221$ was an influential point, our methodology allows us to classify and describe its effect on the model. The application illustrates how our proposed approach identifies known influential observations in a simple and intuitive way. Further, it provides insight into the specific effect of the observations on the model.

\subsection{High-dimensional case: Weight gain after kidney transplants}
\label{subsec:adipose}

Weight gain after kidney transplantation is known to be problematic. Substantial weight gain \citep[according to][averaging at an increase of $12$ kg]{patel1998effect} results in an increased risk of adverse health effects for the transplant patients. As the effect of calorie intake on weight gain is highly individual, genetic variation has also been taken into account. \citet{cashion2013expression} investigated the predictive power of genomic data regarding weight gain, measuring gene expression profiles in subcutaneous adipose tissue of $26$ kidney transplant patients. The tissue samples were collected at the time of surgery, and the mRNA levels were measured using Affymetrix Human Gene 1.0 ST arrays, resulting in gene expression profiles for $28869$ genes\footnote{Data are available in the EMBL-EBI ArrayExpress database (\texttt{www.ebi.ac.uk/arrayexpress}) under accession number E-GEOD-33070.}. The change in weight was recorded after 6 months, which we used to build a predictive model for weight gain based on ridge regression. Our interest lies in identifying potentially influential observations (patients), and their influence on the model complexity of the ridge model. 

\begin{figure}%
  \includegraphics[width=0.49\textwidth]{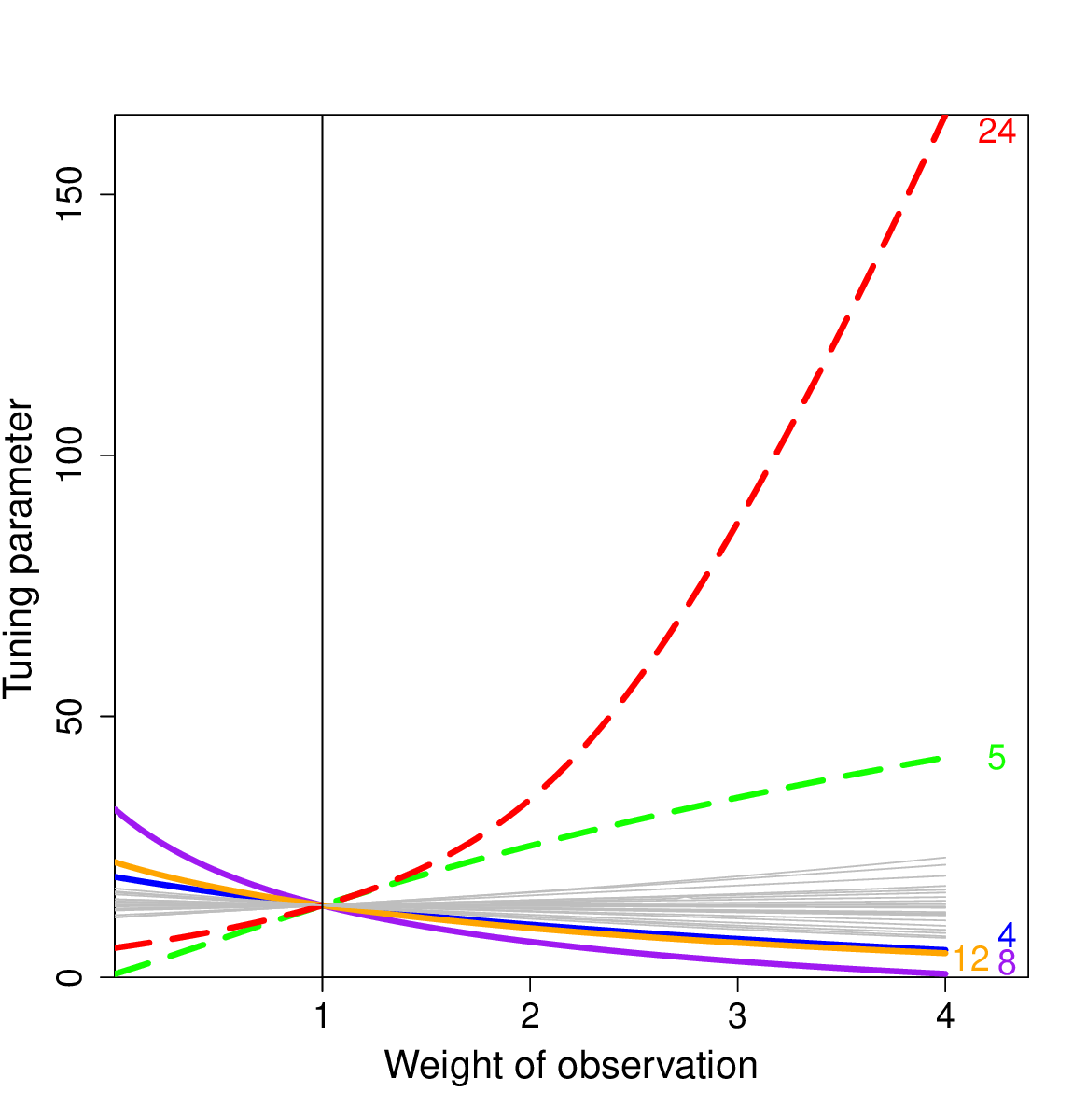}
  \includegraphics[width=0.49\textwidth]{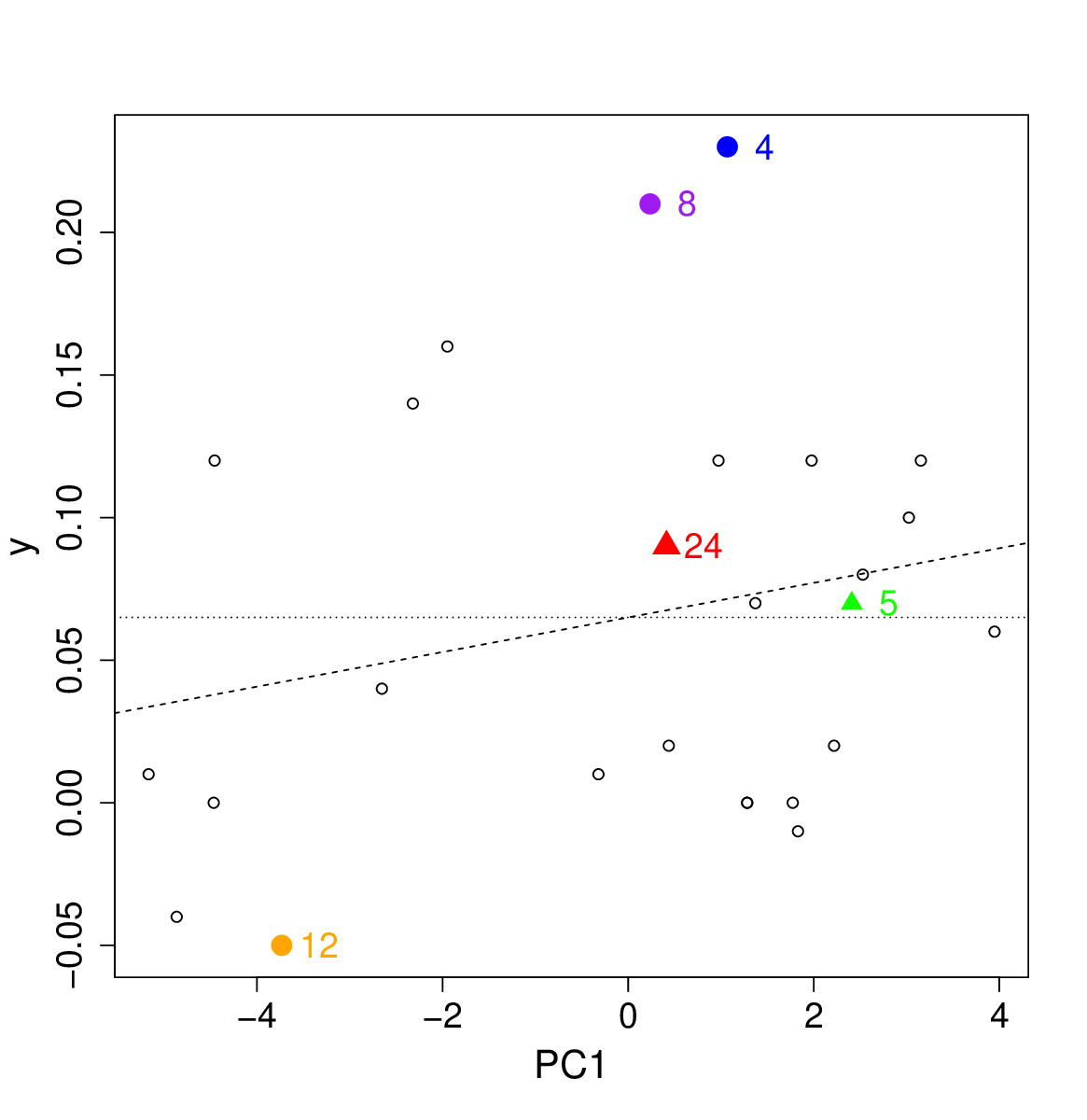}
\caption{Curves of $\hat\lambda_{\cv}(w)$ as functions of $w_i$ for the observations in the kidney transplant dataset and scatter plot of their responses against the first principal component.}
\label{fig:AdiposeScatter}
\end{figure}

The analysis is performed in the same manner as that of Section \ref{sec:bodyfat}, and the covariates are scaled to unit variance. The outcome is the weight gain relative to the initial body weight. Figure \ref{fig:AdiposeScatter} shows the curves of the optimal tuning parameter as a function of the weight, with a scatter plot of the outcome against the first principal component (PC). For this high-dimensional dataset, the first principal component explains $18.51\%$ of the total variation.

\begin{figure}[t]%
\centering
\includegraphics[width=0.49\textwidth]{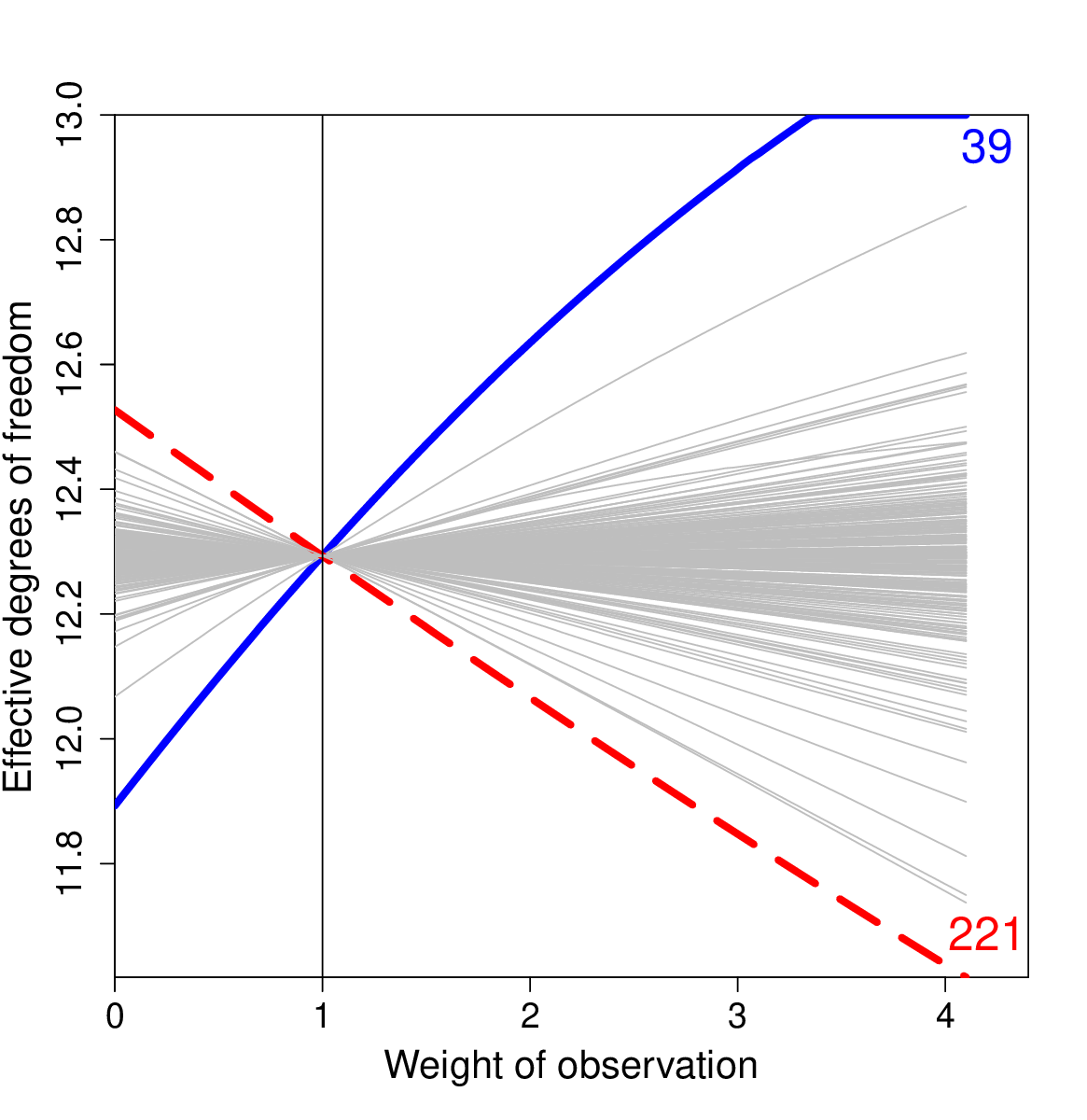}
 \includegraphics[width=0.49\textwidth]{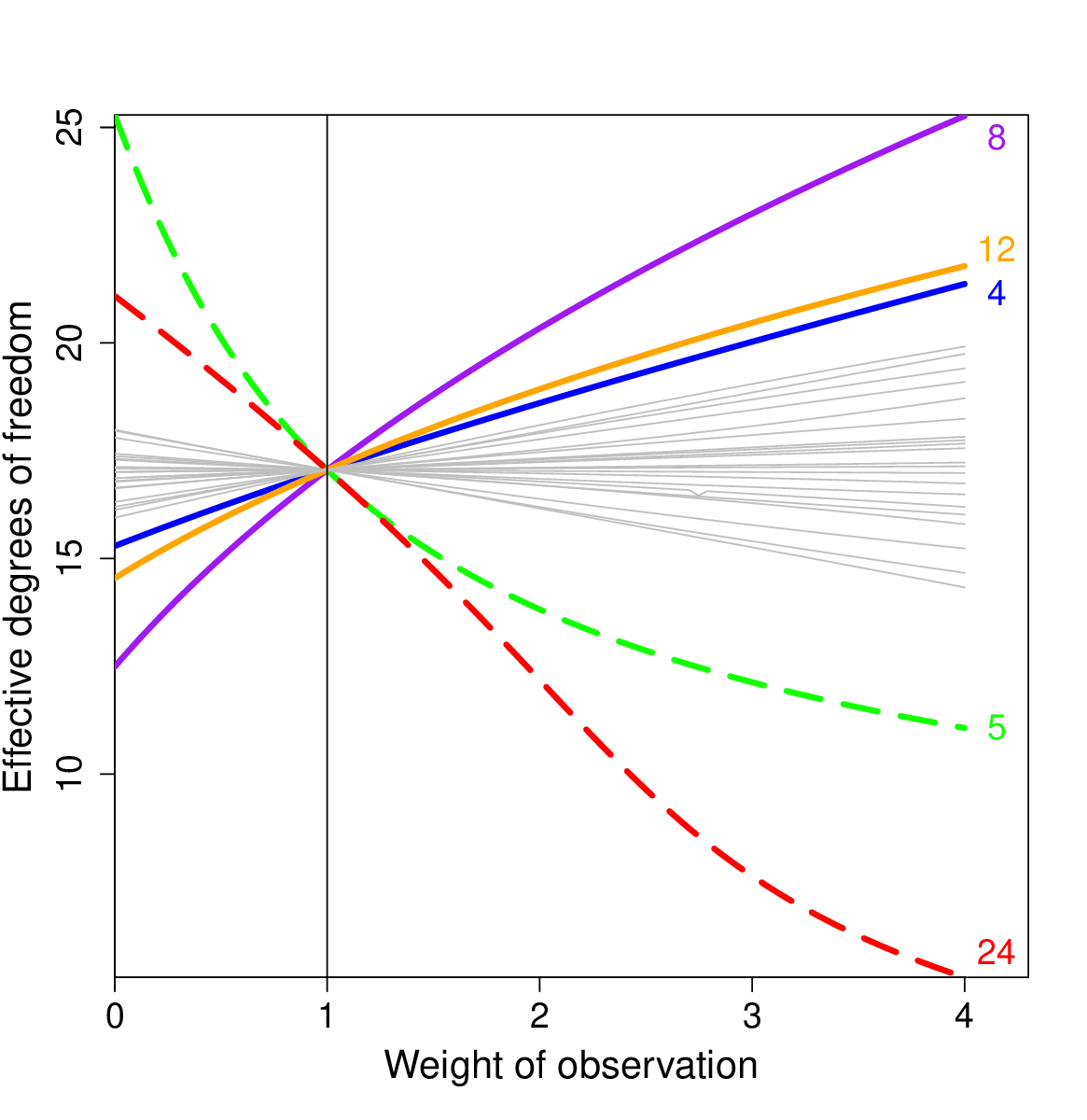}
\caption{Curves of the effective degrees of freedom for the body fat and kidney transplant datasets.}%
\label{fig:ApplicationsEffectiveDF}%
\end{figure}

Observation $24$ stands out as an influential point, a shrinker, with a large increase in $\hat{\lambda}_{\cv}(w_i)$ when given more weight. It has an outcome closer to the mean and by emphasizing this observation a less complicated model is required. Or, equivalently, an increased tuning parameter value. For high-dimensional data, in particular, it may be easier to interpret the change in the effective degrees of freedom rather than the value of the penalty parameter itself, as seen in Section \ref{subsect:shrinkersExpanders}. The right panel of Figure \ref{fig:ApplicationsEffectiveDF} shows the curves of $\hat{\lambda}_{\cv}(w_i)$ on the scale of effective degrees of freedom. It is clear that observation $24$ reduces the model complexity, but its gradient at $w_i=1$ is now not much different from that of observation $5$. Moreover, the effective degrees of freedom scale displays more clearly the effect of observations $4$, $8$ and $12$ as expanders. Observation $8$, in particular, has similar covariate values as observation $24$ but with a large residual, and giving it more weight leads to increased model complexity. The right panel of Figure \ref{fig:AdiposeScatter} shows that observations $8$ and $24$ have similar covariate values in the first PC. As we noted in the simulated data examples (Section \ref{sect:simulations}), the influence of single observations is strongly related to their position with respect to the other points. While observation $4$ is close to observation $8$, its covariate values are slightly farther away from observation $24$ than those of the latter. If we only consider the leverages, we would expect a higher effect for observation $4$, but the relative position to observation $24$ actually reinforces the effect of observation $8$. This may also explain why observation $24$ is such a strong shrinker, as it contrasts the effect of both observations $4$ and $8$.

Observation $5$ is another shrinker whose curve of $\hat{\lambda}_{\cv}(w_i)$ stands out. Its outcome, given the covariate values, is close to the mean, enabling it to counteract the expanding effect of the other observations with similar covariate values but larger outcomes. It is interesting to note that down-weighting this observation has a stronger effect than down-weighting observation $24$. As we have seen in the simulated data example (bottom panel of Figure \ref{fig:Simulations}), the effect of perturbing the weight is not necessarily proportional, and it is important to evaluate the whole curve. In this regard, our method is better than those based on deletion, which only report what happens at weight $0$. In this example, the effect of observation $5$ would have been incorrectly classified as stronger than that of observation $24$ if a deletion method had been used.

This example shows how the methodology of this paper easily generalizes to higher dimensions, unlike other methods such as the one of \citet{DeBin17Influential}. Some caution should be exercised in the interpretation of the curves of the tuning parameter and we would instead advocate the use of effective degrees of freedom. Note that for low-dimensional data this is not required. When we contrast the left plot of Figure \ref{fig:ApplicationsEffectiveDF} with that of Figure \ref{fig:BodyfatScatter}, we do not see any significant difference, except for the reversed ordering. 

\subsection{Comparison with \cite{RajaratnamAl2019}}

When we compare our method with influence-lasso \citep{RajaratnamAl2019} on the body fat dataset (Section \ref{sec:bodyfat}), we obtain a result in line with that of the simulated data examples: influence-lasso identifies more observations than our approach as influential (namely observations 38, 82, 108, 171, 221, 224, 3, 86, 207, 42 and 39, ordered by increasing influence). While observations 39 and 221 are in common with our results, many other points are found as more influential than observation 221. This may be a consequence of the fact that observation 221 does not have neither high leverage nor high residuals (it mainly affects the choice of the tuning parameter by contrasting the effect of observation 39), so it may not have an high value in any of the four measures considered by influence-lasso.

Probably for the same reason, influence-lasso does not identify many of the influential points our method found in the kidney transplant dataset (Section \ref{subsec:adipose}). There influence-lasso only identified observation 4, which has a very large residual, as influential. On the other hand, our method also found observations 5, 8, 12 and 24, despite they do not have neither large residuals nor high leverage.

\section{Discussion}\label{sect:discussion}
We studied the effect of single observations on the cross-validation-based choice of the ridge penalty parameter. We identified two different types of possible influence, one that increases the model complexity (performed by points we termed ``expanders'') and one that reduces it (by points similarly termed ``shrinkers''). Our differentiation approach, based on a continuous perturbation of the weights, improves the traditional methods based on deletion and provides better insight into the effect of the observations. \cite{DeBin17Influential} already pointed in this direction by considering multiple inclusion of single observations in a bootstrap sample, but could not evaluate the effect of an observation when it was down-weighted but not completely excluded. At the same time, due to the lack of bootstrap samples including a single observation several times, \cite{DeBin17Influential} could not evaluate the effect of a single observation when its weight is strongly increased. Our approach also allows for that: the Supplementary material contains an additional example in which, partly due to the large sample size ($n = 120$), there are no clear influential points (Figure S.4, top panels). If we still wish to analyze the effect of a single observation, we may increase the weights as seen in the bottom panels of Figure S.4. On the one hand, it may make little sense to investigate something that occurs only when a single observation is replicated several (let us say $10$) times. On the other hand, the weight should be related to the number of other observations, and a large weight is necessary to see the effect of the characteristics of an observation in a large sample.

While our approach resembles resampling-based approaches, we would like to point out that, in contrast to the latter, it does not require a repeated application of a procedure on several (pseudo-)samples. The advantages in terms of speed are noticeable. Moreover, in contrast to several existing methods, our approach scales well to high-dimensional data. Initial investigations into other penalized regression methods, such as the lasso, indicates that the concepts of expanders and shrinkers are still valid, but with non-smooth or non-differentiable tuning parameter curves. As for all cases in which a close-form expression for the cross-validation error is not available, one has to rely on numerical methods to compute its derivative \citep[but for some special cases, see][Supplementary material]{RajaratnamAl2019}. While the loss in speed is not significant, it is more difficult to interpret the results.

In Section \ref{sect:ridge} we stated that, in practice, the penalty parameter is often chosen via 5- or 10-fold cross validation, while our approach is has been presented based on leave-one-out cross validation. The latter allows us to directly investigate the influence of a single observation on the model (complexity), but one may be interested in the influence of the single observations within the actual cross-validation version used in the tuning process. While approximations for the results of formulas \eqref{cv_res} and \eqref{cv} are available for the generic $K$-fold cross validation \citep{meijer2013efficient}, it is not possible to separate the effect of the single observations within the fold, unless the cross-validation procedure is repeated several times. Therefore, either the results are presented at a fold level, or a (computational) time consuming resampling approach should be implemented. While the latter lacks one of the desirable feature of our approach (speed), it may give additional interesting insights into the problem (e.g., combined influence of pairs of observations) and will be investigated in the future.

As seen in the simulated data (Section \ref{sect:simulations}) and in the educational body fat (Section \ref{sect:realData}) examples, our method identifies the known outliers as highly influential points. Therefore our method can also be seen as an exploratory tool for their identification. An observation that, when up-weighted, requires a much larger tuning parameters (strong shrinker), indeed, is likely to be generated from a different (simpler) mechanism than the rest. Similarly, an observation that, when up-weighted, requires a much smaller tuning parameter (a strong expander) is likely to be generated from a more complex mechanism than that of the other observations. It is worth noting, however, that the strength of shrinkers and expanders is not necessarily connected to the chance of an observation being an outlier. For example, a ``benevolent'' outlier, i.e., one that does not alter the model despite being separated from the rest \citep[see][Ch.\ 2, in particular Fig.\ 2.1(c)]{BelsleyAl1980}, will probably be only a weak shrinker/expander, while a perfectly ``average'' observation may be a strong shrinker just by contrasting the effect of an outlier, as was the case for observation 221 in the Body Fat example. Issues may also arise due to the so-called ``masking effect'' \citep{BendreKale1985}: if the effect of an observation is hidden by that of a stronger one, its influence will be evident only for large values of the weight, i.e., on the right tail of the $\hat{\lambda}_{CV}(w)$ curves. In this case, it is particularly important to look at the entire curve and not limit the attention to the most informative point ($w_i = 1/n, \forall i$). All in all, our graphical approach may be seen as a quick and useful diagnostic tool to decide whether specific methods for outliers detection \citep[and, possibly, robust inference tools, see, e.g.,][]{MaronnaAl2019} should be implemented.

Finally, we preferred to not add to the plots any formal statistical test or threshold delimiting an area in which an ``influential point'' should be considered strong ``ènough'' to compromise the analysis. We prefer to leave this evaluation to the sensibility of the researcher. The danger that such a delicate aspect is reduced to a yes/no decision based on a specific cut-off is, in our opinion, too big. There is the risk, indeed, that such cut-off will be misused as it often happens, for example, for the $\alpha = 0.05$ threshold in variable selection \citep{WassersteinAl2019}. Moreover, as seen in Section \ref{subsec:simulComp}, test-based approaches may find many false positive due to the type I error, especially in the case of large sample sizes. A pure graphical inspection of the $\hat{\lambda}(w_i)$ curves plot should be therefore preferred, the same way the quantile-quantile plot is used to evaluate normality of the residuals in a linear regression model.

\section*{Acknowledgements}
The authors thank Tonje G. Lien for the initial discussions and Mette Langaas for useful comments. The authors have declared no conflict of interest.

\bigskip
\begin{center}
{\large\bf SUPPLEMENTARY MATERIAL}
\end{center}

\begin{description}
\item[Additional results.] Detailed derivations and proofs for Sec.\ \ref{sect:ridge} and \ref{sec:InfluentialMethodology}. Additional figures for the simulations in Sec.\ \ref{sect:simulations} and another high-dimensional data example (.pdf file)
\item[Code.] R code to reproduce all the analyses (.zip file)
\end{description}

\bibliographystyle{chicago}
\bibliography{BibRidge}

\end{document}